\documentclass[aps,prd,showpacs,notitlepage,nofootinbib,preprintnumbers,amsmath,amssymb]{revtex4-1}

\allowdisplaybreaks

\usepackage{graphics,graphicx}
\usepackage{dcolumn}
\usepackage{bm}
\usepackage{epsfig}
\usepackage[usenames]{color}
\usepackage[colorlinks=true,citecolor=blue,linkcolor=blue,urlcolor=blue]{hyperref}
\usepackage{mathbbol}
\usepackage{epstopdf}
\usepackage{simplewick}
\usepackage[utf8]{inputenc} 
\usepackage{slashed}
\usepackage{stackrel}
\usepackage{mathrsfs}
\usepackage{xcolor}
\usepackage{subcaption}
\usepackage[most]{tcolorbox}

\usepackage{hyperref, empheq}
\usepackage[offset=1.5em]{simpler-wick}
\usepackage{xcolor, graphicx, slashed,  accents,  multirow,caption, subcaption}
\usepackage[most]{tcolorbox}

\usepackage{ragged2e} 
\DeclareCaptionJustification{justified}{\justifying}

\def\eq#1{{Eq.~(\ref{#1})}}
\def\eqs#1{{Eqs.~(\ref{#1})}}

\newcommand{\ben}{\begin{eqnarray*}}
\newcommand{\een}{\end{eqnarray*}}
\newcommand{\un}[1]{\underline{#1}}

\newcommand{\og}{\omega \gamma}

\newcommand{\as}{\alpha_s}

\newcommand{\xo}{x_{10}}

\newcommand{\xti}{x_{21}}
\newcommand{\dprime}{\prime \prime}

\def\eq#1{{Eq.~(\ref{#1})}}
\def\eqs#1{{Eqs.~(\ref{#1})}}

\newcommand{\bas}{{\bar{\alpha}_s}}

\newcommand{\dhd}{{\textstyle d}
\lower.03ex\hbox{\kern-0.38em$^{\scriptstyle-}$}\kern-0.05em{}}
\newcommand{\dbar}{{\textstyle \delta}
\lower.03ex\hbox{\kern-0.38em$^{\scriptstyle-}$}\kern-0.05em{}}

\makeatletter
\DeclareRobustCommand{\cev}[1]{%
  {\mathpalette\do@cev{#1}}%
}
\newcommand{\do@cev}[2]{%
  \vbox{\offinterlineskip
    \sbox\z@{$\m@th#1 x$}%
    \ialign{##\cr
      \hidewidth\reflectbox{$\m@th#1\vec{}\mkern4mu$}\hidewidth\cr
      \noalign{\kern-\ht\z@}
      $\m@th#1#2$\cr
    }%
  }%
}
\makeatother

\newcommand{\iUV}{
    \int \displaylimits^z_{\frac{1}{s x_{10}^2}} \frac{dz^\prime }{z^\prime} \int \displaylimits^{x_{10}^2}_{\frac{1}{z^\prime s}} \frac{dx_{21}^2}{x_{21}^2}
}
\newcommand{\iIR}{
\int \displaylimits^{z}_{\frac{\Lambda^2}{s}} \frac{dz^\prime}{z^\prime} \int \displaylimits^{\mathrm{min}[\frac{z}{z^\prime} x_{10}^2, \frac{1}{\Lambda^2}]}_{\mathrm{max}[x_{10}^2, \frac{1}{z^{\prime}s}]} \frac{d x_{21}^2}{x_{21}^2}
}

\newcommand{\inUV}{\int \displaylimits^{z^\prime}_{\frac{1}{sx_{10}^2}} \frac{dz^{\dprime}}{z^{\dprime}} \int \displaylimits^{\mathrm{min}[x_{10}^2, x_{21}^2 \frac{z^\prime}{z^{\dprime}}]}_{\frac{1}{z^{\dprime} s}} \frac{d x_{32}^2}{x_{32}^2}}

\newcommand{\inIR}{\int \displaylimits^{z^\prime \frac{x_{21}^2}{x_{10}^2}}_{\frac{\Lambda^2}{s}} \frac{dz^{\dprime}}{z^{\dprime}} 
    \int \displaylimits^{\mathrm{min}[\frac{z^\prime}{z^{\dprime}} x_{21}^2, \frac{1}{\Lambda^2}]}_{\mathrm{max}[x_{10}^2, \frac{1}{z^{\dprime}s}]} \frac{d x_{32}^2}{x_{32}^2} }

\newcommand{\wb}{\omega_b}

\newcommand{\iUVs}{
    \int \displaylimits^\eta_{s_{10}} d \eta^\prime \int \displaylimits^{\eta^\prime}_{s_{10}} d s_{21}
}
\newcommand{\iIRs}{
    \int \displaylimits^{s_{10}}_{0} d s_{21} \int \displaylimits^{\eta - s_{10} + s_{21}}_{s_{21}} d \eta^\prime
} 

\newcommand{\inIRs}{
    \int \displaylimits^{s_{10}}_0 d s_{32} \int \displaylimits^{\eta^\prime - s_{21} + s_{32}}_{s_{32}} d \eta^{\dprime}
}

\newcommand{\inUVs}{
    \int \displaylimits^{\eta^\prime}_{s_{10}} d \eta^{\dprime} \int \displaylimits^{\eta^{\dprime}}_{\mathrm{max}[s_{10}, s_{21} + \eta^{\dprime} - \eta^\prime]} d s_{32}
}

\newcommand{\iw}{\int\frac{d\omega}{2\pi i}}
\newcommand{\ig}{\int\frac{d\gamma}{2\pi i}}
\newcommand{\ithree}{\bar{I}_{3\omega \gamma}}

\newcommand{\ithreeO}{\bar{I}^{(0)}_{3\omega \gamma}}

\begin{document}

\title{Orbital Angular Momentum Small-$x$ Evolution: Exact Results in the Large-$N_c$ Limit} 

\author{Brandon Manley}
  \email[Email: ]{manley.329@osu.edu}
	\affiliation{Department of Physics, The Ohio State
           University, Columbus, OH 43210, USA}

\begin{abstract}
 We construct an exact solution to the revised small-$x$ orbital angular momentum (OAM) evolution equations derived in \cite{Kovchegov:2023yzd}, based on an earlier work \cite{Kovchegov:2019rrz}. These equations are derived in the double logarithmic approximation (summing powers of $\as \ln^2(1/x)$ with $\as$ the strong coupling constant and $x$ the Bjorken $x$ variable) and the large-$N_c$ limit, with $N_c$ the number of quark colors. From our solution, we extract the small-$x$, large-$N_c$ expressions of the quark and gluon OAM distributions. Additionally, we determine the large-$N_c$ small-$x$ asymptotics of the OAM distributions to be
 \begin{align}
    \notag
     L_{q+\bar{q}}(x,Q^2) \sim L_G(x,Q^2) \sim \Delta \Sigma (x,Q^2) \sim \Delta G(x,Q^2) \sim \left(\frac{1}{x} \right)^{\alpha_h},
 \end{align}
 with the intercept $\alpha_h$ the same as obtained in the small-$x$ helicity evolution \cite{Borden:2023ugd}, which can be approximated as $\alpha_h \approx 3.66074 \sqrt{\frac{\as N_c}{2\pi}}$. This result is in complete agreement with \cite{Boussarie:2019icw}. Additionally, we calculate the ratio of the quark and gluon OAM distributions to the flavor-singlet quark and gluon helicity parton distribution functions respectively in the small-$x$ region. 
\end{abstract}

\maketitle
\tableofcontents


\section{Introduction}
One of the most important open questions in hadronic physics is the proton spin puzzle \cite{Aidala:2012mv, Accardi:2012qut, Leader:2013jra, Aschenauer:2013woa, Aschenauer:2015eha, Boer:2011fh, Proceedings:2020eah, Ji:2020ena, AbdulKhalek:2021gbh}. The proton spin puzzle is best described by spin sum rules, due to Ji \cite{Ji:1996ek} and Jaffe and Manohar \cite{Jaffe:1989jz}. The latter reads 
\begin{align}
    S_{q+\bar{q}} + L_{q+\bar{q}} + S_G + L_G = \frac{1}{2},
\end{align}
where $S_{q+\bar{q}}$ ($S_G$) is the quark (gluon) helicity contribution to the proton spin, and $L_{q+\bar{q}}$ ($L_G$) is the quark (gluon) contribution to the proton spin from the orbital angular momentum (OAM). The reader may find reviews of the proton spin puzzle in \cite{Accardi:2012qut, Leader:2013jra, Aschenauer:2013woa, Aschenauer:2015eha, Proceedings:2020eah, Ji:2020ena}. 

The helicity contributions can be expressed as integrals over the longitudinal momentum fraction $x$ 
\begin{align}
    S_{q+\bar{q}}(Q^2) = \frac{1}{2} \int \displaylimits^1_0 dx\, \Delta \Sigma(x,Q^2), \hspace{1cm} S_G(Q^2) = \int \displaylimits^1_0 dx\, \Delta G(x, Q^2),
\end{align}
where $Q^2$ is the renormalization scale and the flavor-singlet quark helicity parton distribution function (hPDF) is 
\begin{align}
    \Delta \Sigma(x,Q^2) = \sum_{f=u,d,s,\ldots} \left[\Delta q_f(x,Q^2) + \Delta \bar{q}_f (x,Q^2) \right].
\end{align}
$\Delta q_f(x, Q^2)$ and $\Delta \bar{q}_f(x,Q^2)$ are respectively the quark and anti-quark hPDFs of a given flavor $f$. $\Delta G(x,Q^2)$ is the gluon hPDF. The current values of the spin carried by the quarks and gluons as extracted from experimental data are $S_{q+\bar{q}}(Q^2 = 10 \, \mathrm{GeV}^2) = 0.15 \div 0.20$ for $x\in [0.001, 1]$ and $S_G(Q^2 = 10\, \mathrm{GeV}^2) = 0.13 \div 0.26$ for $x\in [0.05, 1]$ \cite{Accardi:2012qut, Leader:2013jra, Aschenauer:2013woa, Aschenauer:2015eha, Proceedings:2020eah, Ji:2020ena}. The fact that these two numbers sum to less than $1/2$ is the proton spin puzzle. The remaining spin could reside at lower values of $x$ and/or the OAM contributions. 
\newline
\indent The OAM contributions can also be written as integrals over the longitudinal momentum fraction $x$ \cite{Bashinsky:1998if, Hagler:1998kg, Harindranath:1998ve, Hatta:2012cs, Ji:2012ba}\footnote{Although we work with the canonical OAM distributions here, see \cite{Banu:2021cla} for translating between the Ji and Jaffe-Manohar definitions specifically at small $x$.},
\begin{align}
    L_{q+\bar{q}}(Q^2) = \int \displaylimits^1_0 dx\, L_{q+\bar{q}}(x,Q^2), \hspace{1cm}L_G(Q^2) = \int \displaylimits^1_0 dx\, L_G(x,Q^2).
\end{align}

Both the helicity PDFs and OAM distributions receive contributions from the small $x$ region. This region is important to study because of the limited amount of data as well as the finite acceptance in $x$ in any given experiment. Since $x$ scales with the inverse of the center-of-mass energy squared, any experiment, even future experiments at the Electron-Ion Collider \cite{Accardi:2012qut, Boer:2011fh, Proceedings:2020eah, AbdulKhalek:2021gbh}, can only probe down to some $x_{\mathrm{min}} > 0$. Therefore, theoretical input is needed to constrain the net amount of spin for values of Bjorken $x < x_{\mathrm{min}}$. \newline
\indent The helicity PDFs, $\Delta \Sigma(x,Q^2)$ and $\Delta G(x,Q^2)$, have been well studied in recent years; they have been experimentally extracted in the accessible region of $x$ and $Q^2$ \cite{Gluck:2000dy, Leader:2005ci, deFlorian:2009vb, Leader:2010rb, Jimenez-Delgado:2013boa, Ball:2013lla, Nocera:2014gqa, deFlorian:2014yva, Leader:2014uua, Sato:2016tuz, Ethier:2017zbq, DeFlorian:2019xxt, Borsa:2020lsz, Zhou:2022wzm, Cocuzza:2022jye}. The OAM contributions have been studied much less. Even though their evolution in $Q^2$ is known \cite{Hagler:1998kg,Hoodbhoy:1998yb}, and there have been some proposals on how to access these distributions experimentally \cite{Bhattacharya:2022vvo, Bhattacharya:2023hbq}, there is currently no experimental extraction of $L_{q+\bar{q}}(x,Q^2)$ and $L_G(x,Q^2)$. In order to fully understand the proton spin puzzle, more work is needed to understand the OAM distributions. As mentioned above, the small $x$ region is particularly important for study. \newline
\indent In the helicity sector, a pioneering work by Bartels, Ermolaev, and Ryskin (BER) \cite{Bartels:1995iu,Bartels:1996wc} employed the infrared evolution equations (IREE) formalism \cite{Gorshkov:1966ht,Kirschner:1983di,Kirschner:1994rq,Kirschner:1994vc,Blumlein:1995jp,Griffiths:1999dj} to study the hPDFs at small-$x$. These evolution equations were derived in the double-logarithmic approximation (DLA), which resums powers of $\as \ln^2(1/x)$, with $\as$ the strong coupling constant. The BER version of the IREE formalism was extended in \cite{Boussarie:2019icw} to study the OAM distributions at small-$x$. \newline
\indent Over the past decade, a new approach to helicity-dependent small-$x$ evolution has been developed in \cite{Kovchegov:2015pbl, Hatta:2016aoc, Kovchegov:2016zex, Kovchegov:2016weo, Kovchegov:2017jxc, Kovchegov:2017lsr, Kovchegov:2018znm, Cougoulic:2019aja, Kovchegov:2020hgb, Cougoulic:2020tbc, Chirilli:2021lif, Kovchegov:2021lvz, Cougoulic:2022gbk}. This new approach is based on the $s$-channel/shock wave formalism previously developed for unpolarized scattering \cite{Mueller:1994rr,Mueller:1994jq,Mueller:1995gb,Balitsky:1995ub,Balitsky:1998ya,Kovchegov:1999yj,Kovchegov:1999ua,Jalilian-Marian:1997dw,Jalilian-Marian:1997gr,Weigert:2000gi,Iancu:2001ad,Iancu:2000hn,Ferreiro:2001qy}. In unpolarized scattering, which dominates at high energies, the evolution in \cite{Mueller:1994rr,Mueller:1994jq,Mueller:1995gb,Balitsky:1995ub,Balitsky:1998ya,Kovchegov:1999yj,Kovchegov:1999ua,Jalilian-Marian:1997dw,Jalilian-Marian:1997gr,Weigert:2000gi,Iancu:2001ad,Iancu:2000hn,Ferreiro:2001qy} is expressed in terms of infinite Wilson lines along the light-cone. This regime is known as the eikonal approximation. Helicity-dependent scattering is suppressed by one power of center-of-mass energy, and is known as the sub-eikonal approximation. This new approach, called the light-cone operator treatment (LCOT) in \cite{Cougoulic:2022gbk}, utilizes sub-eikonal operators inserted between semi-infinite and finite light-cone Wilson lines. \newline
\indent Using the LCOT formalism, small $x$ evolution equations were derived for the polarized dipole amplitudes, which are related to the hPDFs, in \cite{Kovchegov:2015pbl, Kovchegov:2016zex, Kovchegov:2017lsr, Kovchegov:2018znm} (KPS). The KPS equations were derived in the DLA. Recently, these equations received important corrections. It was discovered that additional sub-eikonal operators mix with the ones studied in the original KPS papers under evolution, and thus resulted in revised evolution equations \cite{Cougoulic:2022gbk} (KPS-CTT) (see \cite{Hatta:2016aoc} also). These revised evolution equations for the polarized dipole amplitudes were solved numerically in the large-$N_c$ \cite{Cougoulic:2022gbk}, and large-$N_c\&N_f$ \cite{Adamiak:2023okq} limits (where $N_f$ is the number of quark flavors) as well as analytically in the large-$N_c$ limit \cite{Borden:2023ugd}. Additionally, the large-$N_c\&N_f$ solution has been used for a successful analysis of the polarized deep inelastic scattering (DIS) and semi-inclusive DIS (SIDIS) small-$x$ data \cite{Adamiak:2023yhz}. \newline 
\indent The small-$x$ asymptotics of the hPDFs have been calculated by BER \cite{Bartels:1996wc}, and by solving the KPS-CTT equations \cite{Cougoulic:2022gbk,Borden:2023ugd,Adamiak:2023okq}. It was shown that there is less than a 1\% difference between the results for the intercept (power of $x$) in the large-$N_c$ limit and less than a 3\% difference between the results in the large-$N_c\&N_f$ limit. The origin of the discrepancy is speculated on in the Appendix of \cite{Borden:2023ugd} (see also \cite{Kovchegov:2016zex}). (Additionally, see \cite{Moch:2014sna, Blumlein:2021ryt, Blumlein:2022gpp} for a discrepancy due to scheme dependence between the IREE and the small-$x$ limit of the exact 3-loop calculations of spin-dependent DGLAP anomalous dimensions.)

 The OAM distributions have also been studied in the LCOT framework. An analysis based on the KPS equations was done in \cite{Kovchegov:2019rrz}, but did not include the corrections found in \cite{Cougoulic:2022gbk}. In \cite{Kovchegov:2023yzd}, these corrections were accounted for and the OAM distributions were expressed in terms of the polarized dipole amplitudes and their first impact-parameter moments. These impact-parameter moments were dubbed the ``moment amplitudes'' in \cite{Kovchegov:2023yzd}. Novel evolution equations for the moment amplitudes based on the KPS-CTT equations were constructed and solved numerically in \cite{Kovchegov:2023yzd}. In the previous study based on the KPS equations, the small-$x$ asymptotics of the OAM distributions were significantly different from each other. In the large-$N_c$ limit, more than a 50\% difference was found between the intercepts for the quark and gluon OAM distributions \cite{Kovchegov:2019rrz}. This result was consistent with the discrepancies between the helicity small $x$ asymptotics obtained via the KPS equations and those obtained by BER. After including the corrections which resulted in the KPS-CTT evolution, it was found that the small $x$ asymptotics of the OAM distributions agree with the result from \cite{Boussarie:2019icw}, at least within the precision of the numerical solution in \cite{Kovchegov:2023yzd}. Furthermore, the ratios of the OAM distributions to the hPDFs were studied in \cite{Kovchegov:2023yzd} and compared to the results in \cite{Boussarie:2019icw}. The ratios in the quark and gluon ratios were found to be in good numerical agreement, with a small discrepancy reported in the gluon sector.

In this paper, we seek to elucidate the numerical findings of \cite{Kovchegov:2023yzd} by solving the large-$N_c$ evolution equations for the moment amplitudes, \eq{all_oam_eqns} below, analytically. In Section \ref{largeNcEqns}, we recall these evolution equations and the relations between the OAM distributions and the polarized dipole amplitudes and their first impact-parameter moments. We present the solution of the moment amplitude evolution equations in Section \ref{sec:sol}, which is based on the double Laplace transform method of \cite{Borden:2023ugd}. A summary of our solution and the resulting small-$x$ asymptotics of the quark and gluon OAM distributons is presented in Section \ref{sec:summary}. Importantly, we find the OAM distributions have the same asymptotics as the hPDFs and the $g_1$ structure function, given in \eq{oam_asymp} as well as in the Abstract above. We trace the origin of this homogeneity to the mixing of the polarized dipole amplitudes with the moment amplitudes in the moment evolution equations. \newline
\indent In Section \ref{sec:ratios}, we use the solution of the moment amplitudes and the KPS-CTT equations to calculate the ratios of the OAM distributions to the hPDFs in the quark and gluon sectors analytically. We confirm the numerical results of  \cite{Kovchegov:2023yzd} and speculate on the small discrepancy with the results in \cite{Boussarie:2019icw}. We conclude in Section \ref{sec:conclusion}.

\section{Evolution equations for the moment amplitudes in the large-$N_c$ limit}
\label{largeNcEqns}

The large-$N_c$ DLA OAM evolution equations, as  derived in \cite{Kovchegov:2023yzd}, evolve the first impact-parameter moments of the polarized dipole amplitudes $G_{10}(zs)$ and $G^i_{10}(zs)$. The operator definitions of the polarized dipole amplitudes, given in \cite{Cougoulic:2022gbk}, are in terms of infinite light-cone Wilson lines and polarized light-cone Wilson lines. The latter are expressed as sub-eikonal operators inserted between semi-infinite light-cone Wilson lines. The large-$N_c$ DLA helicity evolution equations involve the impact-parameter integrated polarized dipole amplitudes, $G(x_{10}^2, zs)$ and $G_2(x_{10}^2, zs)$ defined by 
\begin{subequations}
    \label{helicity_defs}
\begin{align}
    \int d^2 x_1 G_{10}(zs) &= G(x_{10}^2, zs), \\ 
    \int d^2 x_1 G^i_{10}(zs) &= x_{10}^i G_1(x_{10}^2,zs) + \epsilon^{ij} x_{10}^j G_2(x_{10}^2, zs).
\end{align}
\end{subequations}
As we will see below, $G_1(x_{10}^2,zs)$ does not contribute to the helicity PDFs or the evolution of $G(x_{10}^2,zs)$ and $G_2(x_{10}^2,zs)$. We use $\un x = (x^1, x^2)$ to denote two-dimensional transverse vectors, and $\un x_{ij} = \un x_i - \un x_j$ for $i,j=0,1,2,\ldots$ labeling the partons. The amplitudes above depend on $x_{10}^2 = |\un x_{10}|^2$, the transverse size squared of the dipole. Additionally, the amplitudes depend on the center of mass energy between the original projectile and the target, $s$, multiplied by the smallest momentum fraction of the two partons making up the dipole, $z$.\footnote{$z$ is more aptly thought of as the parameter controlling the evolution of $zs$ \cite{Kovchegov:2015pbl, Kovchegov:2018znm, Kovchegov:2021lvz}. For instance, after a step of evolution involving a virtual correction, $z$ could be smaller than either of the two longitudinal momentum fractions of the partons making up the dipole.} The OAM distributions depend not only on $G(x_{10}^2,zs)$ and $G_2(x_{10}^2,zs)$, but also on the first impact-parameter moments of $G_{10}(zs)$ and $G^i_{10}(zs)$ defined as 
\begin{subequations}
    \label{moment_defs}
\begin{align}
    \label{i3_eqn}
    \int d^2 x_1 \, x_1^i G_{10}(zs) &= \xo^i I_3(\xo^2, zs) + \cdots,
    \\ \label{i4_eqn}
    \int d^2 x_1 \,x_1^i G^j_{10}(zs) &= \epsilon^{ij} \xo^2 
    I_{4}(\xo^2, zs) + \epsilon^{ik} \xo^k \xo^j I_{5}(\xo^2, zs) + \epsilon^{jk} \xo^k \xo^i I_{6}(\xo^2, zs) + \cdots, 
\end{align}
\end{subequations}
where the ellipses denote additional tensor structures that do not contribute to the OAM distributions. The amplitudes $I_3, I_4, I_5$ and $I_6$ were dubbed the ``moment" amplitudes in \cite{Kovchegov:2023yzd}. The evolution of the moment amplitudes mixes them with $G(x_{10}^2, zs)$ and $G_2(x_{10}^2 ,zs)$ and was derived in \cite{Kovchegov:2023yzd}.

In addition to the amplitudes in \eqs{helicity_defs} and (\ref{moment_defs}), the evolution for the moment amplitudes (as well the helicity evolution) depends on the so-called neighbor dipole amplitudes, denoted by $\Gamma_{10,21}(zs)$ and $\Gamma^i_{10,21}(zs)$, which are auxiliary amplitudes necessary to enforce lifetime ordering in the evolution \cite{Kovchegov:2015pbl}. Their operator definitions are the same as those for $G_{10}(zs)$ and $G^i_{10}(zs)$, except for a difference in the light-cone lifetime cutoff, which depends on the adjacent dipole size \cite{Kovchegov:2018znm, Cougoulic:2019aja, Cougoulic:2022gbk}. One can write decompositions analogous to \eqs{helicity_defs} and (\ref{moment_defs}) for the neighbor amplitudes,
\begin{subequations}
\begin{align}
    \int d^2 x_1 \, \Gamma_{10,21}(zs) &=  \Gamma(\xo^2, x_{21}^2, zs), \\ 
    \int d^2 x_1 \,  \Gamma^i_{10,21}(zs) &= \epsilon^{ij} x_{10}^j \Gamma_2(\xo^2, x_{21}^2, zs) + \cdots ,
    \\
    \int d^2 x_1 \, x_1^i \Gamma_{10,21}(zs) &= \xo^i \Gamma_3(\xo^2, x_{21}^2, zs) + \cdots,
    \\
    \label{gammai_decomp}
    \int d^2 x_1 \,x_1^i \Gamma^j_{10,21}(zs) &= \epsilon^{ij} \xo^2 
    \Gamma_4(\xo^2, x_{21}^2, zs) + \epsilon^{ik} \xo^k \xo^j \Gamma_5(\xo^2, x_{21}^2, zs) + \epsilon^{jk} \xo^k \xo^i \Gamma_6(\xo^2, x_{21}^2, zs) + \cdots, 
\end{align}
\end{subequations}
where again the ellipses denote additional tensor structures that do not contribute to the helicity PDFs, OAM distributions or the evolution of the dipole and moment amplitudes. Neither the helicity PDFs nor the OAM distributions depend on the neighbor amplitudes directly. The neighbor amplitudes only contribute to the evolution of the dipole and moment amplitudes. As we will see below, the helicity PDFs and OAM distributions depend only on the polarized dipole and moment amplitudes. 

The DLA evolution equations for the moment amplitudes in the large-$N_c$ limit are \cite{Kovchegov:2023yzd}

\begin{align} \label{all_oam_eqns}
    \begin{pmatrix}
    I_3 \\ 
    I_4 \\
    I_5 \\
    I_6 
    \end{pmatrix}(\xo^2, zs) &= \begin{pmatrix}
    I_3^{(0)} \\ 
    I_4^{(0)} \\
    I_5^{(0)} \\
    I_6^{(0)} 
    \end{pmatrix} (\xo^2, zs)  \\ \notag 
    & \hspace{1cm}
    + \frac{\as N_c}{4\pi} \iUV \,  \begin{pmatrix}
    2\,\Gamma_3 - 4\,\Gamma_4 + 2 \, \Gamma_5 + 6\, \Gamma_6 - 2\, \Gamma_2 \\ 
    0 \\
    0 \\
    0 
    \end{pmatrix}(\xo^2, \xti^2, z^\prime s) 
    \\\notag & \hspace{1cm}
    + \frac{\as N_c}{4\pi} \iIR \, \begin{pmatrix}
        4 & -4 & 2 & 6 & -4 &  - 6 \\
        0 & 4 & 2 & -2 & 0 &   1 \\
        -2 & 2 & -1 & -3 & 2 &  3 \\
        0 & 0 & 0 & 0 & 2 &  4 \\
    \end{pmatrix} 
    \begin{pmatrix}
    I_3 \\ 
    I_4 \\
    I_5 \\
    I_6 \\ 
    G \\
    G_2
    \end{pmatrix} (\xti^2, z^\prime s),
\end{align}
where the inhomogeneous terms are the initial conditions of the evolution. The DLA evolution equations for the neighbor moment amplitudes are \cite{Kovchegov:2023yzd}
\begin{align}
    \begin{pmatrix} \label{oam_neigh_eqn}
    \Gamma_3 \\ 
    \Gamma_4 \\
    \Gamma_5 \\
    \Gamma_6 
    \end{pmatrix}(\xo^2, \xti^2, z^\prime s) &= \begin{pmatrix}
    I_3^{(0)} \\ 
    I_4^{(0)} \\
    I_5^{(0)} \\
    I_6^{(0)} 
    \end{pmatrix} (\xo^2, z^\prime s)  \\ \notag 
    & \hspace{1cm}
    + \frac{\as N_c}{4\pi} \inUV \,  \begin{pmatrix}
    2\,\Gamma_3 - 4\,\Gamma_4 + 2 \, \Gamma_5 + 6\, \Gamma_6 - 2\, \Gamma_2 \\ 
    0 \\
    0 \\
    0 
    \end{pmatrix}(\xo^2, x_{32}^2, z^{\dprime} s) 
    \\\notag &\hspace{1cm}
    + \frac{\as N_c}{4\pi} \inIR 
    \, \begin{pmatrix}
        4 & -4 & 2 & 6 & -4 &  -6 \\
        0 & 4 & 2 & -2 & 0 &  1 \\
        -2 & 2 & -1 & -3 & 2 &  3 \\
        0 & 0 & 0 & 0 & 2 &  4 \\
    \end{pmatrix} 
    \begin{pmatrix}
    I_3 \\ 
    I_4 \\
    I_5 \\
    I_6 \\ 
    G \\
    G_2
    \end{pmatrix} (x_{32}^2, z^{\dprime} s),
\end{align}
where $\Gamma_p(x_{10}^2, x_{21}^2, z^\prime s)$ for $p=2,3,4,5,6$ are only defined for $x_{10} \geq x_{21}$, and $1/\Lambda$ is an infrared (IR) cutoff for all dipole sizes. 

The dipole amplitudes $G$ and $G_2$, in addition to the neighbor amplitude $\Gamma_2$, appear explicitly in \eqs{all_oam_eqns} and (\ref{oam_neigh_eqn}). Therefore, to solve \eqs{all_oam_eqns} and (\ref{oam_neigh_eqn}), one needs to first solve the large-$N_c$ helicity evolution equations, Eqs.~(133) of \cite{Cougoulic:2022gbk}, to obtain $G, G_2,$ and $\Gamma_2$. This has been done in \cite{Borden:2023ugd}. We will use their solution for $G, G_2, \Gamma_2$ below to solve \eqs{all_oam_eqns} and (\ref{oam_neigh_eqn}). 

For brevity, we will use the following linear combination of amplitudes
\begin{subequations}
    \label{shiftedI3}
    \begin{align}
        \bar{I}_3(x_{10}^2,zs) &\equiv I_3(x_{10}^2, zs) + 2 \, I_5(x_{10}^2, zs), \\ 
        \bar{\Gamma}_3(x_{10}^2, x_{21}^2, z^\prime s) &\equiv \Gamma_3(x_{10}^2, x_{21}^2, z^\prime s) + 2\, \Gamma_5(x_{10}^2, x_{21}^2, z^\prime s),
    \end{align}
\end{subequations}
in place of $I_3(x_{10}^2, zs)$ and $\Gamma_3(x_{10}^2 ,x_{21}^2, z^\prime s)$ in \eqs{all_oam_eqns} and (\ref{oam_neigh_eqn}). 
Additionally, to simplify the algebra and notation, we introduce vectorial notation for the moment amplitudes appearing in \eqs{i4_eqn} and (\ref{gammai_decomp})
\begin{subequations}
    \label{vector_not}
    \begin{align}
        \label{i_vec}
        \vec{I}(x_{10}^2, zs) &\equiv \Big(I_4(x_{10}^2,zs), \, I_5(x_{10}^2,zs), \, I_6(x_{10}^2,zs)  \Big), \\ 
         \Vec{\Gamma}(x_{10}^2,x_{21}^2, z^\prime s), &\equiv \Big(\Gamma_4(x_{10}^2,x_{21}^2, z^\prime s), \,\Gamma_5(x_{10}^2,x_{21}^2, z^\prime s), \,\Gamma_6(x_{10}^2,x_{21}^2, z^\prime s)  \Big).
    \end{align}
\end{subequations}
We may also express the initial conditions for the amplitudes in \eq{i_vec} in a similar vector notation
\begin{align}
    \vec{I}^{(0)}(x_{10}^2, zs) &\equiv \Big(I^{(0)}_4(x_{10}^2,zs), \, I^{(0)}_5(x_{10}^2,zs), \, I^{(0)}_6(x_{10}^2,zs)  \Big).
\end{align}
Finally, it is more useful to work in terms of the rescaled variables \cite{Kovchegov:2016weo, Kovchegov:2020hgb, Cougoulic:2022gbk}
\begin{subequations}
    \label{scaled_variables}
\begin{align}
    \eta =  \sqrt{\frac{\as N_c}{2\pi}} \ln \frac{zs}{\Lambda^2},
    \hspace{0.5cm}
    \eta^\prime  &=  \sqrt{\frac{\as N_c}{2\pi}} \ln \frac{z^\prime s}{\Lambda^2},
    \hspace{0.5cm}
    \eta^{\dprime} =  \sqrt{\frac{\as N_c}{2\pi}} \ln \frac{z^{\dprime}s}{\Lambda^2},
    \\
    s_{10} =  \sqrt{\frac{\as N_c}{2\pi}} \ln \frac{1}{x_{10}^2 \Lambda^2},
    \hspace{0.5cm}
    s_{21} &=  \sqrt{\frac{\as N_c}{2\pi}} \ln \frac{1}{x_{21}^2 \Lambda^2}, 
    \hspace{0.5cm}
    s_{32} =  \sqrt{\frac{\as N_c}{2\pi}} \ln \frac{1}{x_{32}^2 \Lambda^2}.
\end{align}
\end{subequations}

Utilizing \eqs{shiftedI3}-(\ref{scaled_variables}) in \eqs{all_oam_eqns} and (\ref{oam_neigh_eqn}), we may write the moment amplitude and neighbor evolution equations as 
\begin{subequations}
    \label{vec_oam_eqns}
\begin{align}
    \label{bari3_eqn}
    \bar{I}_3(s_{10}, \eta) &= \bar{I}_3^{(0)}(s_{10}, \eta) +\iUVs \Big[\,\bar{\Gamma}_3(s_{10}, s_{21}, \eta^\prime) -  \Gamma_2(s_{10}, s_{21}, \eta^\prime)  
    + \vec{V}_{\mathrm{UV}} \cdot \vec{\Gamma}(s_{10}, s_{21}, \eta^\prime)  \Big], 
    \\ 
    \label{sep_vec_eqn}
   \vec{I}(s_{10}, \eta) &= \vec{I}^{(0)} (s_{10}, \eta)  
    + \frac{1}{2} \iIRs \, 
    \Big[ \bar{I}_3(s_{21}, \eta^\prime) \vec{V}_{\bar{I}_3} + G(s_{21}, \eta^\prime) \vec{V}_G 
    \\ \notag
    &
    \hspace{8cm}
    + G_2(s_{21}, \eta^\prime) \vec{V}_{G_2} + M_{\mathrm{IR}} \vec{I}(s_{21}, \eta^\prime) \Big],
    \\
    \label{gamma3_eqn}
    \bar{\Gamma}_3(s_{10}, s_{21}, \eta^\prime) &= \bar{I}_3^{(0)}(s_{10}, \eta^\prime) + \inUVs \Big[\bar{\Gamma}_3 (s_{10}, s_{32}, \eta^{\dprime}) - \Gamma_2(s_{10}, s_{32}, \eta^{\dprime}) 
    \\ \notag
    & \hspace{10cm}
    + \vec{V}_{\mathrm{UV}} \cdot \vec{\Gamma}(s_{10}, s_{32}, \eta^{\dprime}) \Big], \\ 
    \label{sep_neigh_eqn}
    \vec{\Gamma}(s_{10}, s_{21}, \eta^\prime) &= \vec{I}^{(0)} (s_{10}, \eta^\prime) 
    + \frac{1}{2} \inIRs
    \, \Big[ \bar{I}_3(s_{32}, \eta^{\dprime}) \vec{V}_{\bar{I}_3} + G(s_{32}, \eta^{\dprime}) \vec{V}_G 
    \\ \notag
    &
    \hspace{8cm}
    + G_2(s_{32}, \eta^{\dprime}) \vec{V}_{G_2} + M_{\mathrm{IR}} \vec{I}(s_{32}, \eta^{\dprime}) \Big],
\end{align}
\end{subequations}
where $0 \leq s_{10} \leq s_{21} \leq \eta^\prime$, and we have changed the order of integration in the third lines of \eqs{all_oam_eqns} and (\ref{oam_neigh_eqn}). Additionally, we have defined 
\begin{subequations}
\begin{align}
    \vec{V}_{\mathrm{UV}} &\equiv (-2, -1,3), \\
    \vec{V}_{\bar{I}_3} &\equiv (0, -2, 0), \\
    \vec{V}_{G} &\equiv (0, 2, 2), \\
    \vec{V}_{G_2} &\equiv (1, 3, 4), \\
    M_{\mathrm{IR}} &\equiv 
    \begin{pmatrix}
         4 & 2 & -2  \\
         2 & -1 & -3  \\
         0 & 0 & 0 \\
    \end{pmatrix} .
\end{align}
\end{subequations}

The task is now to solve \eqs{vec_oam_eqns}. Furthermore, since the moment equations do not close on their own as mentioned above, we will need the solution for $G, G_2$ and $\Gamma_2$ from \cite{Borden:2023ugd}. This solution is expressed in double Laplace transforms, and reads (see Eqs. (53) and (54) of \cite{Borden:2023ugd})
\begin{subequations}
    \label{helsol_1}
\begin{align}
    \label{g2_sol}
    G_2(s_{10}, \eta) &= \iw\ig \, e^{\omega( \eta- s_{10}) + \gamma s_{10}} G_{2 \omega \gamma}, \\ 
   G(s_{10}, \eta) &= \iw\ig \, e^{\omega( \eta - s_{10}) + \gamma s_{10}} G_{\og}, \\ 
    \label{gm2_sol}
   \Gamma_2(s_{10}, s_{21}, \eta^\prime) &= \iw\ig \,
   \left[
    e^{\omega(\eta^\prime - s_{21}) + \gamma s_{10}}\left( G_{2 \omega \gamma} - G_{2 \omega \gamma}^{(0)} \right) + e^{\omega (\eta^\prime - s_{10}) + \gamma s_{10}} G_{2\omega \gamma}^{(0)}
   \right],
\end{align}
\end{subequations}
where the double Laplace images are 
\begin{subequations}
    \label{helsol_2}
\begin{align}
 \label{g2wg_sol}
    G_{2\omega \gamma} &= G^{(0)}_{2 \omega \gamma} + \frac{2 (\gamma - \delta^+_\omega) \left( G^{(0)}_{\delta^+_\omega \gamma} + 2 G^{(0)}_{2 \delta^+_\omega \gamma} \right) - 2 (\gamma_\omega^+ - \delta^+_\omega) \left( G^{(0)}_{\delta^+_\omega \gamma^+_\omega} + 2 G^{(0)}_{2 \delta^+_\omega \gamma_\omega^+} \right) + 8 \,\delta_\omega^- \left( G^{(0)}_{2 \omega \gamma} - G_{2 \omega \gamma^+_\omega}^{(0)}\right) }{\omega (\gamma - \gamma_\omega^-)(\gamma - \gamma_\omega^+)},
    \\
    G_{\og} &= \frac{1}{2} \og (G_{2\og} - G^{(0)}_{2\og}) - 2 G_{2\og}.
\end{align}
\end{subequations}
Here, the initial conditions of the helicity evolution have also been expressed in terms of double Laplace transforms
\begin{subequations}
    \label{helsol_ics}
    \begin{align}
    G^{(0)}(s_{10}, \eta) &= \iw\ig \, e^{\omega(\eta - s_{10}) + \gamma s_{10}} G_{\omega \gamma}^{(0)}, \\ 
    G^{(0)}_2(s_{10}, \eta) &= \iw\ig \, e^{\omega(\eta - s_{10}) + \gamma s_{10}} G_{2\omega \gamma}^{(0)},
\end{align}
\end{subequations}
and, following \cite{Borden:2023ugd}, we have also defined
\begin{subequations}
    \begin{align}
        \label{del_def}
        \delta_\omega^{\pm} &\equiv \frac{\omega}{2} \left[ 1 \pm \sqrt{1 - \frac{4}{\omega^2}} \right], \\ 
        \gamma_\omega^{\pm} &\equiv \frac{\omega}{2} 
        \left[
            1 \pm \sqrt{1 - \frac{16}{\omega^2} \sqrt{1- \frac{4}{\omega^2}}}
        \right].
    \end{align}
\end{subequations}
As usual, the contours in \eqs{helsol_1} and (\ref{helsol_ics}) run parallel to the imaginary axis to the right of all the singularities of the integrands. Note, for the contours in \eqs{helsol_1} and (\ref{helsol_ics}), $\mathrm{Re}\,\omega > \mathrm{Re} \,\gamma$ \cite{Borden:2023ugd}.

The solution of \eqs{vec_oam_eqns}, along with \eqs{helsol_1}, will give us both the canonical OAM distributions and helicity PDFs at small-$x$ and large-$N_c$, by employing the following relations derived in \cite{Cougoulic:2022gbk, Kovchegov:2023yzd, Kovchegov:2019rrz} 
\begin{subequations}
    \label{dist_rels}
    \begin{align}
        \label{quarkOAMrel}
        L_{q+\bar{q}}(x,Q^2) &= \frac{N_f}{\as \pi^2} 
    \int \displaylimits^{\sqrt{\bas} \ln \frac{Q^2}{ \Lambda^2}}_0 d s_{10}
    \int \displaylimits^{s_{10} + \sqrt{\bas} \ln \frac{1}{x}}_{s_{10}} d \eta \Big[ G(s_{10}, \eta) {+} 3\, G_2(s_{10}, \eta) 
        - \bar{I}_3(s_{10}, \eta) - \vec{V}_{\mathrm{UV}}  \cdot \vec{I}(s_{10}, \eta)
        \Big],  
        \\ 
         L_G(x,Q^2) &= - \frac{2 \, N_c}{\as \pi^2} \Big(2\, \vec{V}_G + V_{\bar{I}_3} -\vec{V}_{\mathrm{UV}} \Big) \cdot \vec{I} \left(s_{10} = \sqrt{\bas} \ln \frac{Q^2}{\Lambda^2}, \eta = \sqrt{\bas} \ln \frac{Q^2}{x\Lambda^2 } \right), 
    \\ \label{quarkHelrel}
        \Delta \Sigma(x,Q^2) &= -\frac{N_f}{\as \pi^2} 
    { \int \displaylimits^{\sqrt{\bas} \ln \frac{Q^2}{ \Lambda^2}}_0 d s_{10}
    \int \displaylimits^{s_{10} + \sqrt{\bas} \ln \frac{1}{x}}_{s_{10}} d \eta}
    \Big[ G(s_{10}, \eta) +  2\, G_2(s_{10}, \eta) 
        \Big], 
    \\ 
        \Delta G(x,Q^2) &= \frac{2 \, N_c}{\as \pi^2} G_2 \left(s_{10} = \sqrt{\bas} \ln \frac{Q^2}{\Lambda^2}, \eta = \sqrt{\bas} \ln \frac{Q^2}{x\Lambda^2 }\right),
    \end{align}
\end{subequations}
where we have defined $\bas = \frac{\as N_c}{2\pi}$. Here we have assumed all flavors contribute equally so that summing over flavors results in a factor of $N_f$ in \eqs{quarkOAMrel} and (\ref{quarkHelrel}). This assumption will need to be revised for phenomenology, where $G$ and $I_3$ would depend on flavor (see \cite{Adamiak:2023yhz} for example). 
Note we have neglected the derivatives from Eq.~(36) of \cite{Kovchegov:2023yzd}, and Eq.~(42) of \cite{Cougoulic:2022gbk}. Such derivatives remove one logarithm of energy and are therefore outside of our DLA approximation.

\section{Solution of the moment amplitude evolution equations}
\label{sec:sol}

Inspired by the success of the double Laplace transform method used to solve the helicity evolution equations \cite{Borden:2023ugd} (see also \cite{Kovchegov:2017jxc}), we also use double inverse Laplace representations here to solve \eqs{vec_oam_eqns}. To start, let us write the moment amplitudes, $\bar{I}_3(s_{10}, \eta), \vec{I}(s_{10}, \eta)$, and their inhomogeneous terms, $\bar{I}_3^{(0)}(s_{10}, \eta), \vec{I}^{(0)}(s_{10}, \eta)$, as

\begin{subequations}
    \label{dm_ansatze}
    \begin{align}
        \label{dm_I_reps}
        \bar{I}_3 (s_{10}, \eta) &= \iw\ig e^{\omega(\eta - s_{10}) + \gamma s_{10}} \bar{I}_{3 \omega \gamma}, \\ 
        \bar{I}^{(0)}_3 (s_{10}, \eta) &= \iw\ig e^{\omega(\eta - s_{10}) + \gamma s_{10}} \bar{I}^{(0)}_{3 \omega \gamma}, \\
        \vec{I} (s_{10}, \eta) &= \iw\ig e^{\omega(\eta - s_{10}) + \gamma s_{10}} \vec{I}_{ \omega \gamma}, \\ 
        \vec{I}^{(0)} (s_{10}, \eta) &= \iw\ig e^{\omega(\eta - s_{10}) + \gamma s_{10}} \vec{I}^{(0)}_{\omega \gamma}.
    \end{align}
\end{subequations}
As mentioned above, we take the contours to the right of all the singularities of the integrands in \eqs{dm_ansatze}. Using \eqs{dm_ansatze}, we will solve \eqs{sep_vec_eqn} and (\ref{sep_neigh_eqn}) first. To start, we can relate the different double Laplace images in \eqs{dm_ansatze} by using them in \eq{sep_vec_eqn}. Doing the $s_{21}$ and $\eta^\prime$ integrals, we get
\begin{align}
    \label{i4_eqn1}
    \iw\ig e^{\omega(\eta-s_{10}) + \gamma s_{10}} 
    \vec{I}_{\omega \gamma}
     &=  \iw\ig \frac{1}{\omega \gamma} e^{\omega(\eta - s_{10}) + \gamma s_{10}}  \vec{I}_{\omega \gamma}^{(0)}
     \\  \notag
     & \hspace{-3cm} + \frac{1}{2}\iw\ig \frac{1}{\omega \gamma} \left( e^{\omega( \eta - s_{10}) + \gamma s_{10}} -  e^{\gamma s_{10}} -  e^{\omega( \eta - s_{10})} + 1\right)
     \Big(
        \bar{I}_{3\omega\gamma} \vec{V}_{\bar{I}_3} + G_{\og} \vec{V}_G + G_{2\og} \vec{V}_{G_2} + M_{\mathrm{IR}} \vec{I}_{\omega \gamma}
     \Big).
\end{align}
Treating $\eta-s_{10}$ and $s_{10}$ as separate variables, we can perform the forward Laplace transforms in \eq{i4_eqn1} to see 
\begin{align}
    \label{i4i5_eqn}
    \vec{I}_{\og} 
    &= \vec{I}^{(0)}_{\og} + \frac{1}{2\,\omega \gamma} 
    \Big(
        \bar{I}_{3\og} \vec{V}_{\bar{I}_3} + G_{\og} \vec{V}_{G} + G_{2\og} \vec{V}_{G_2} + M_{\mathrm{IR}} \vec{I}_{\og}
    \Big),
\end{align}
or, solving for $\vec{I}_{\og}$,
\begin{align}
    \vec{I}_{\og} =(2 \og - M_{\mathrm{IR}})^{-1} \Big(2 \og \,\vec{I}^{(0)}_{\og} + G_{\og} \vec{V}_{G} + G_{2\og} \vec{V}_{G_2} + \vec{V}_{\bar{I}_3} \bar{I}_{3\og} \Big).
\end{align}

Next, we observe the following scaling relation from \eqs{sep_vec_eqn} and (\ref{sep_neigh_eqn}),
\begin{align}
    \label{gamma_scaling}
    \vec{\Gamma} (s_{10}, s_{21}, \eta^\prime) - \vec{I}^{(0)}(s_{10}, \eta^\prime) = \vec{I}(s_{10}, \eta = \eta^\prime + s_{10} - s_{21}) - \vec{I}^{(0)}(s_{10}, \eta = \eta^\prime + s_{10} - s_{21}),
\end{align}
which, after using \eqs{dm_ansatze}, gives   
\begin{align}
\label{gmI_solns}
    \vec{\Gamma}(s_{10}, s_{21}, \eta^\prime) &= \iw\ig \left[ 
     e^{\omega(\eta^\prime -s_{21}) + \gamma s_{10}} \left( \vec{I}_{ \omega \gamma} -  \vec{I}^{(0)}_{\omega \gamma} \right) + e^{\omega(\eta^\prime -s_{10}) + \gamma s_{10}} \vec{I}_{\omega \gamma}^{(0)} 
    \right].
\end{align}

\eqs{sep_vec_eqn} and (\ref{sep_neigh_eqn}) have several boundary conditions that need to be satisfied. We explicitly check that our solution satisfies these boundary conditions in Appendix \ref{sec:BC}. Therefore, we conclude \eqs{sep_vec_eqn} and (\ref{sep_neigh_eqn}) are completely solved. 

Let us now solve \eqs{bari3_eqn} and (\ref{gamma3_eqn}). One can show, after differentiating \eq{gamma3_eqn}, that $\bar{\Gamma}_3$ obeys the following second-order partial differential equation
\begin{align}
    \label{gm3_pde}
    \Bigg[ \frac{\partial^2}{\partial s_{21}^2} + \frac{\partial^2}{\partial s_{21} \partial \eta^\prime} + 1
    \Bigg] \bar{\Gamma}_3 (s_{10}, s_{21}, \eta^\prime) &= - \vec{V}_{\mathrm{UV}} \cdot \vec{\Gamma}(s_{10}, s_{21}, \eta^\prime) + \Gamma_2(s_{10}, s_{21},\eta^\prime).
\end{align}

This differential equation has a homogeneous and a particular solution, which we label $\bar{\Gamma}_3^{(h)}$ and $\bar{\Gamma}_3^{(p)}$ respectively. The general solution of \eq{gm3_pde} would then be a sum of $\bar{\Gamma}_3^{(h)}$ and $\bar{\Gamma}_3^{(p)}$. 
Searching for a homogeneous solution of the form 
\begin{align}
    \bar{\Gamma}_3^{(h)}(s_{10}, s_{21}, \eta^\prime) = \int \frac{d\omega}{2\pi i} \, e^{\omega (\eta^\prime - s_{21}) + \gamma s_{21}} \bar{\Gamma}_{3\og}(s_{10})
\end{align}
yields the condition
\begin{align}
    \label{hom_cond}
    \gamma^2 - \og + 1 = 0. 
\end{align}
The two solutions to \eq{hom_cond} are $\gamma = \delta^+_\omega$ and $\gamma = \delta^-_\omega$ as defined above in \eq{del_def}. The complete homogeneous solution to \eq{gm3_pde} is then 
\begin{align}
    \label{hom_res}
        \bar{\Gamma}_3^{(h)}(s_{10}, s_{21}, \eta^\prime) = \int \frac{d\omega}{2\pi i} \, e^{\omega (\eta^\prime - s_{21})} \Big[ e^{\delta^+_\omega s_{21}} \bar{\Gamma}_{3\omega}^+(s_{10}) + e^{\delta^-_\omega s_{21}} \bar{\Gamma}_{3\omega}^-(s_{10})        \Big],
\end{align}
where $\bar{\Gamma}^\pm_{3\omega}(s_{10})$ are arbitrary functions of $s_{10}$ that can be constrained by boundary conditions. We explicitly determine these functions in Appendix \ref{sec:constraints}; they are listed in \eq{appenB_gmres} and below in \eq{gm_funcs}.

Using \eqs{gm2_sol} and (\ref{gmI_solns}) in the right hand side of \eq{gm3_pde}, one can easily show the particular solution is 
\begin{align}
    \label{par_res}
    &\bar{\Gamma}^{(p)}_3(s_{10}, s_{21}, \eta^\prime) = \iw \ig \Bigg\{ e^{\omega(\eta^\prime - s_{10}) + \gamma s_{10}} \left[G_{2\og}^{(0)} - \vec{V}_{\mathrm{UV}} \cdot \vec{I}^{(0)}_{\og} \right] 
    \\ \notag 
    & \hspace{6cm} + e^{\omega(\eta^\prime- s_{21}) + \gamma s_{10}}
    \Big[ G_{2\og} - G^{(0)}_{2\og} -\vec{V}_{\mathrm{UV}} \cdot \Big(\vec{I}_{\og} - \vec{I}^{(0)}_{\og} \Big) \Big]
    \Bigg\}.
\end{align}
Via \eqs{hom_res} and (\ref{par_res}), the general solution to \eq{gm3_pde} is 
\begin{align}
    \label{gm3_res}
    \bar{\Gamma}_3(s_{10}, s_{21}, \eta^\prime) &= \iw e^{\omega(\eta^\prime -s_{21})} \left[\bar{\Gamma}^+_{3\omega}(s_{10}) \, e^{\delta^+_\omega s_{21}} + \bar{\Gamma}^-_{3\omega}(s_{10}) \, e^{\delta^-_\omega s_{21}} \right] 
    \\ \notag 
    & + \iw \ig \Bigg\{ e^{\omega(\eta^\prime - s_{10}) + \gamma s_{10}} \left[ G_{2\og}^{(0)} - \vec{V}_{\mathrm{UV}} \cdot \vec{I}^{(0)}_{\og} \right] 
    \\ \notag 
    & \hspace{4cm} + e^{\omega(\eta^\prime- s_{21}) + \gamma s_{10}}
   \Big[ G_{2\og} - G^{(0)}_{2\og}  -\vec{V}_{\mathrm{UV}} \cdot \Big(\vec{I}_{\og} - \vec{I}^{(0)}_{\og} \Big)\Big]
    \Bigg\}.
\end{align}

To ensure consistency with the solution in \eq{helsol_1}, we also choose $\mathrm{Re}\,\omega > \mathrm{Re}\,\gamma$ here. As mentioned above, the functions $\bar{\Gamma}^\pm_{3\omega}(s_{10})$ and the image $\bar{I}_{3\og}$ can be constrained via boundary conditions from \eqs{bari3_eqn} and (\ref{gamma3_eqn}). This is done in Appendix \ref{sec:constraints}, and we simply quote the results here

\begin{subequations}
    \label{appen_res}
\begin{align}
    \label{gm_funcs}
    \bar{\Gamma}^{\pm}_{3\omega}(s_{10}) &= 
    e^{- \delta^{\pm}_\omega s_{10}} \frac{\delta^\pm_{\omega}}{\delta^\pm_\omega - \delta^\mp_\omega} \ig e^{\gamma s_{10}}
    \Bigg\{(\omega \delta^\mp_\omega -1) \Big[ G_{2\og} - G_{2\og}^{(0)} - \vec{V}_{\mathrm{UV}} \cdot \left(\vec{I}_{\og} - \vec{I}_{\og}^{(0)} \right) \Big] 
    \\ \notag 
    & \hspace{10cm}
    - G_{2\og}^{(0)} + \vec{V}_{\mathrm{UV}} \cdot \vec{I}^{(0)}_{\og} + \ithree
    \Bigg\},
    \\ 
    \label{i3_exp}
    \bar{I}_{3\og} &= 
    \frac{\vec{V}_{\mathrm{UV}} \cdot \Big(\vec{I}^{(0)}_{\og} - \vec{I}^{(0)}_{\delta^+_\omega \gamma} \Big)- G^{(0)}_{2\og} + G^{(0)}_{2 \delta^+_\omega \gamma} - \bar{I}^{(0)}_{3\delta^+_\omega \gamma}}{(\omega \delta^-_\omega -1) \vec{V}_{\mathrm{UV}} \cdot (2\og - M_{\mathrm{IR}})^{-1} \vec{V}_{\bar{I}_3} -1} 
    \\ \notag 
   & +  \frac{(\omega \delta^-_\omega -1)\Big[
        G_{2\og} - G^{(0)}_{2\og} + \vec{V}_{\mathrm{UV}} \cdot \vec{I}^{(0)}_{\og} - \vec{V}_{\mathrm{UV}} \cdot (2 \og - M_{\mathrm{IR}})^{-1}(2 \og \vec{I}^{(0)}_{\og} + G_{\og} \vec{V}_{G} + G_{2\og} \vec{V}_{G_2})
    \Big] }{(\omega \delta^-_\omega -1) \vec{V}_{\mathrm{UV}} \cdot (2\og - M_{\mathrm{IR}})^{-1} \vec{V}_{\bar{I}_3} -1}.
\end{align}
\end{subequations}
 We have now completely solved \eqs{vec_oam_eqns}. That is, given the initial conditions to the evolution, $\bar{I}^{(0)}_3(s_{10}, \eta), \vec{I}^{(0)}(s_{10}, \eta)$ as well as $G^{0}(s_{10}, \eta)$ and $G_2^{(0)}(s_{10}, \eta)$, we solve \eqs{vec_oam_eqns} via \eqs{dm_ansatze}, (\ref{gmI_solns}), (\ref{i4i5_eqn}), (\ref{gm3_res}), and (\ref{appen_res}).

\section{Summary of our solution and the small-$x$ asymptotics}
\label{sec:summary}

\subsection{Summary of our solution}
In the previous Section, we constructed an exact solution to \eqs{vec_oam_eqns}. We reiterate the solution here for convenience
\begin{subequations}
    \label{the_sol}
\begin{align}
    \bar{I}_3(s_{10}, \eta) &= \ig \iw \, e^{\omega (\eta - s_{10}) + \gamma s_{10}} \ithree , \\ 
    \vec{I}(s_{10}, \eta) &= \ig \iw e^{\omega (\eta - s_{10}) + \gamma s_{10}}  (2 \og - M_{\mathrm{IR}})^{-1} \Big(2 \og \,\vec{I}^{(0)}_{\og} + G_{\og} \vec{V}_{G} + G_{2\og} \vec{V}_{G_2} + \vec{V}_{\bar{I}_3} \bar{I}_{3\og} \Big),
    \\ 
\bar{\Gamma}_3(s_{10}, s_{21}, \eta^\prime) &= \iw e^{\omega(\eta^\prime -s_{21})} \left[\bar{\Gamma}^+_{3\omega}(s_{10}) \, e^{\delta^+_\omega s_{21}} + \bar{\Gamma}^-_{3\omega}(s_{10}) \, e^{\delta^-_\omega s_{21}} \right] 
    \\ \notag 
    & + \iw \ig \Bigg\{ e^{\omega(\eta^\prime - s_{10}) + \gamma s_{10}} \left[- \vec{V}_{\mathrm{UV}} \cdot \vec{I}^{(0)}_{\og} + G_{2\og}^{(0)} \right] 
    \\ \notag 
    & \hspace{-2cm} + e^{\omega(\eta^\prime- s_{21}) + \gamma s_{10}}
   \Big[G_{2\og} - G^{(0)}_{2\og} + \vec{V}_{\mathrm{UV}} \cdot \vec{I}^{(0)}_{\og} - \vec{V}_{\mathrm{UV}} \cdot (2 \og - M_{\mathrm{IR}})^{-1}(2 \og \vec{I}^{(0)}_{\og} + G_{\og} \vec{V}_{G} + G_{2\og} \vec{V}_{G_2}) \Big]
    \Bigg\}
    \\
   \vec{\Gamma}(s_{10}, s_{21}, \eta^\prime) &= \iw\ig \Bigg\{ 
   e^{\omega(\eta^\prime -s_{10}) + \gamma s_{10}} \vec{I}_{\omega \gamma}^{(0)} 
   \\ \notag
   & \hspace{0cm} + e^{\omega(\eta^\prime -s_{21}) + \gamma s_{10}} \left[(2\og - M_{\mathrm{IR}})^{-1} \Big(2 \og \vec{I}^{(0)}_{\og} + \vec{V}_{G} G_{\og} + \vec{V}_{G_2} G_{2\og} + \vec{V}_{\bar{I}_3} \bar{I}_{3\og} \Big) -  \vec{I}^{(0)}_{\omega \gamma} \right] 
    \Bigg\},
\end{align}
\end{subequations}

where 
\begin{subequations}
    \label{aux_defs}
\begin{align}
    f^{(0)}(s_{10}, \eta) &= \iw \ig e^{\omega(\eta-s_{10}) + \gamma s_{10}} f^{(0)}_{\og}, \\
        \bar{I}_{3\og} &= 
    \frac{\vec{V}_{\mathrm{UV}} \cdot \Big(\vec{I}^{(0)}_{\og} - \vec{I}^{(0)}_{\delta^+_\omega \gamma} \Big)- G^{(0)}_{2\og} + G^{(0)}_{2 \delta^+_\omega \gamma} - \bar{I}^{(0)}_{3\delta^+_\omega \gamma}}{(\omega \delta^-_\omega -1) \vec{V}_{\mathrm{UV}} \cdot (2\og - M_{\mathrm{IR}})^{-1} \vec{V}_{\bar{I}_3} -1}
    \\ \notag 
   & +  \frac{(\omega \delta^-_\omega -1)\Big[
        G_{2\og} - G^{(0)}_{2\og} + \vec{V}_{\mathrm{UV}} \cdot \vec{I}^{(0)}_{\og} - \vec{V}_{\mathrm{UV}} \cdot (2 \og - M_{\mathrm{IR}})^{-1}(2 \og \vec{I}^{(0)}_{\og} + G_{\og} \vec{V}_{G} + G_{2\og} \vec{V}_{G_2})
    \Big] }{(\omega \delta^-_\omega -1) \vec{V}_{\mathrm{UV}} \cdot (2\og - M_{\mathrm{IR}})^{-1} \vec{V}_{\bar{I}_3} -1} ,
    \\ 
    \bar{\Gamma}^{\pm}_{3\omega}(s_{10}) &= e^{- \delta^{\pm}_\omega s_{10}} \frac{\delta^\pm_{\omega}}{\delta^\pm_\omega - \delta^\mp_\omega} \ig e^{\gamma s_{10}}
    \Bigg\{(\omega \delta^\mp_\omega -1) \Big[ G_{2\og} - G_{2\og}^{(0)} - \vec{V}_{\mathrm{UV}} \cdot \left(\vec{I}_{\og} - \vec{I}_{\og}^{(0)} \right) \Big]
    \\ \notag 
    & \hspace{10cm}
    - G_{2\og}^{(0)} + \vec{V}_{\mathrm{UV}} \cdot \vec{I}^{(0)}_{\og} + \ithree
    \Bigg\},
    \\
    G_{2\omega \gamma} &= G^{(0)}_{2 \omega \gamma} + \frac{2 (\gamma - \delta^+_\omega) \left( G^{(0)}_{\delta^+_\omega \gamma} + 2 G^{(0)}_{2 \delta^+_\omega \gamma} \right) - 2 (\gamma_\omega^+ - \delta^+_\omega) \left( G^{(0)}_{\delta^+_\omega \gamma^+_\omega} + 2 G^{(0)}_{2 \delta^+_\omega \gamma_\omega^+} \right) + 8 \,\delta_\omega^- \left( G^{(0)}_{2 \omega \gamma} - G_{2 \omega \gamma^+_\omega}^{(0)}\right) }{\omega (\gamma - \gamma_\omega^-)(\gamma - \gamma_\omega^+)},
    \\
    G_{\og} &= \frac{1}{2} \og (G_{2\og} - G_{2\og}^{(0)}) - 2 G_{2\og},
\end{align}
\end{subequations}
for $f \in \{G, G_2, \bar{I}_3, \vec{I} \}$, and, as defined above, we have 
\begin{subequations}
    \begin{align}
    \delta_\omega^{\pm} &\equiv \frac{\omega}{2} \left[ 1 \pm \sqrt{1 - \frac{4}{\omega^2}} \right], \\ 
    \gamma_\omega^{\pm} &\equiv \frac{\omega}{2} 
        \left[
            1 \pm \sqrt{1 - \frac{16}{\omega^2} \sqrt{1- \frac{4}{\omega^2}}}
        \right],
    \\ 
    \vec{V}_{\mathrm{UV}} &\equiv (-2, -1,3), \\
    \vec{V}_{\bar{I}_3} &\equiv (0, -2, 0), \\
    \vec{V}_{G} &\equiv (0, 2, 2), \\
    \vec{V}_{G_2} &\equiv (1, 3, 4), \\
    M_{\mathrm{IR}} &\equiv 
    \begin{pmatrix}
         4 & 2 & -2  \\
         2 & -1 & -3  \\
         0 & 0 & 0 \\
    \end{pmatrix} .
\end{align}
\end{subequations}
 With the exact expressions for the polarized dipole and moment amplitudes in hand, we can now write down the helicity PDFs and OAM distributions. Plugging \eqs{the_sol} with \eq{aux_defs} into \eqs{dist_rels} gives 
\begin{subequations}
    \label{exact_dists}
    \begin{align}
    \label{eLq}
    L_{q+\bar{q}}(x, Q^2) &= \frac{N_f}{\as \pi^2} \iw \ig \frac{e^{\sqrt{\bas} \omega \ln(1/x) + \sqrt{\bas} \gamma \ln(Q^2/\Lambda^2)}}{ {\omega \gamma} } \Big[ 
    {G_{\og} + 3\, G_{2\og} - \bar{I}_{3\og} }
    \\ \notag &
    \hspace{3cm} 
     {- \vec{V}_{\mathrm{UV}} \cdot (2 \og - M_{\mathrm{IR}})^{-1} \Big(2 \og \,\vec{I}^{(0)}_{\og} + G_{\og} \vec{V}_{G} + G_{2\og} \vec{V}_{G_2} + \vec{V}_{\bar{I}_3} \bar{I}_{3\og} \Big) \Big]}, 
    \\ \label{eLG}
    L_G(x, Q^2) &= -\frac{2N_c}{\as \pi^2} \iw \ig e^{\sqrt{\bas} \omega \ln(1/x) + \sqrt{\bas} \gamma \ln(Q^2/\Lambda^2)} 
    \\ \notag 
    & \hspace{2cm} \times 
    \Big( 2\, \vec{V}_{G} -\vec{V}_{\mathrm{UV}} + \vec{V}_{\bar{I}_3} \Big) \cdot (2 \og - M_{\mathrm{IR}})^{-1} \Big(2 \og \,\vec{I}^{(0)}_{\og} + G_{\og} \vec{V}_{G} + G_{2\og} \vec{V}_{G_2} + \vec{V}_{\bar{I}_3} \bar{I}_{3\og} \Big),
    \\ \label{eDE}
    \Delta \Sigma(x, Q^2) &= -\frac{N_f}{\as \pi^2} \iw \ig \frac{e^{\sqrt{\bas} \omega \ln(1/x) + \sqrt{\bas} \gamma \ln(Q^2/\Lambda^2)}}{{\omega\gamma}} \Big[ 
    G_{\og} + 2\, G_{2\og} \Big], 
    \\ \label{eDG}
     \Delta G(x, Q^2) &= \frac{2N_c}{\as \pi^2} \iw \ig e^{\sqrt{\bas} \omega \ln(1/x) + \sqrt{\bas} \gamma \ln(Q^2/\Lambda^2)} 
    G_{2\og},
    \end{align}
\end{subequations}
where, as above in \eqs{dist_rels}, we have defined $\bas \equiv \frac{\as N_c}{2\pi}$. In arriving at \eqs{eLq} and (\ref{eDE}), we have done the $s_{10}$ and $\eta$ integrals in \eqs{quarkOAMrel} and (\ref{quarkHelrel}). We emphasize here that, while \eqs{eLq} and (\ref{eLG}) were derived in \cite{Kovchegov:2023yzd}, here we have explicitly found the double Laplace images. As mentioned above, \eqs{eDE} and (\ref{eDG}) were derived in \cite{Cougoulic:2022gbk} and the double Laplace images were found in \cite{Borden:2023ugd}.  

All together, \eqs{exact_dists} give the exact small-$x$, large-$N_c$ DLA expressions for the quark and gluon OAM distributions and the quark and gluon helicity PDFs. 

\subsection{Small-$x$ asymptotics of the OAM distributions}

From \eqs{eLq} and (\ref{eLG}), we can see that the small-$x$ asymptotics of the OAM distributions are governed by the rightmost singularity in the $\omega$-plane. One can show this singularity is given by setting the large square root in $\gamma_\omega^-$ to $0$ \cite{Borden:2023ugd}, which gives
\begin{align}
    \label{asymp_sing}
    1- \frac{16}{\omega^2} \sqrt{1- \frac{4}{\omega^2}} = 0.
\end{align}
The solution to \eq{asymp_sing} with the largest real part is \cite{Borden:2023ugd}
\begin{align}
    \label{rm_sing}
    \omega_b \equiv \frac{4}{3^{1/3}} \sqrt{\mathrm{Re} \left[\Big(-9 + i \sqrt{111} \Big)^{1/3} \right]} \approx 3.66074.
\end{align}
Via \eqs{eLq} and (\ref{eLG}), the small-$x$ asymptotics of the quark and gluon OAM distributions are then
\begin{align} \label{oam_asymp}
    L_{q+\bar{q}}(x, Q^2) \sim L_G(x,Q^2) \sim \left( \frac{1}{x} \right)^{\omega_b  \sqrt{\frac{\as N_c}{2\pi}}}.
\end{align}
\eq{oam_asymp}, along with the solution of the small-$x$, large-$N_c$ evolution equations for the moment amplitudes, \eqs{the_sol}, represent the main result of this work. 

Importantly, the OAM distributions have the same intercept $\alpha_h \equiv \omega_b \sqrt{\bas}$ as the helicity PDFs. In fact, one can explicitly remove (by hand) the mixing with $G, G_2,$ and $\Gamma_2$ and solve \eqs{vec_oam_eqns}. The resulting solution has small-$x$ asymptotics that are also governed by the rightmost singularity in the $\omega$-plane. In this case however, one finds the rightmost singularity to be $\omega_b^\prime = 2 < \omega_b$. This is similar to the study of the OAM distributions based on the KPS equations in \cite{Kovchegov:2019rrz}. There, the ``moment amplitudes" were sub-leading at small-$x$ and did not mix with $G,G_2,$ or $\Gamma_2$ under evolution. 

Therefore, one can trace the leading small-$x$, large-$N_c$ asymptotic behavior of the OAM distributions to the mixing of the moment amplitudes with the helicity amplitudes $G, G_2,$ and $\Gamma_2$ under evolution.\footnote{Note that $G(s_{10}, \eta)$ and $G_2(s_{10}, \eta)$ also explicitly appear in the expression for the quark OAM distribution, \eq{quarkOAMrel}.} This situation is similar to the polarized DGLAP evolution equations for the OAM distributions in the Wandzura-Wilczek approximation \cite{Hagler:1998kg,Hoodbhoy:1998yb}.  There, the (twist-2 part of the) OAM distributions mix with the helicity PDFs. The leading asymptotic behavior of the (twist-2 part of the) OAM distributions is given by the anomalous dimensions of the helicity PDFs. As we do not employ the Wandzura-Wilczek approximation in this work, it appears the inclusion of genuine twist-3 terms, at least in the case of the small-$x$ evolution presented here, does not affect this conclusion. 

We should compare the result in \eq{oam_asymp} to the one obtained in \cite{Boussarie:2019icw}. In \cite{Boussarie:2019icw}, the BER IREE formalism was extended to the OAM distributions. The resulting OAM distributions were found to obey the same small-$x$ asymptotics as the helicity PDFs. Namely, they found (see Eq. (7) of \cite{Boussarie:2019icw})

\begin{align} \label{OAMBERintercept}
    L_{q+\bar{q}}(x,Q^2) \sim L_G(x,Q^2) \sim \left(\frac{1}{x}\right)^{\alpha^{BER}_h},
\end{align}
where $\alpha^{BER}_h$ is the intercept of both the quark and gluon helicity PDFs. The analytic expression for $\alpha^{BER}_h$ in the large-$N_c$ limit, given in \cite{Kovchegov:2016zex} and for which the detailed solution is given in \cite{Borden:2023ugd}, is 
\begin{align}
     \alpha^{BER}_h \equiv \sqrt{\frac{17 + \sqrt{97}}{2}} \sqrt{\frac{\as N_c}{2\pi}} \approx 3.66394 \sqrt{\frac{\as N_c}{2\pi}}.
\end{align}

Comparing \eqs{oam_asymp} and (\ref{OAMBERintercept}), we see a slight difference between the intercepts of the OAM distributions found above in the LCOT formalism and found in \cite{Boussarie:2019icw} using the BER IREE formalism. This small discrepancy is the same as the one between the helicity PDF intercepts obtained in the LCOT formalism \cite{Cougoulic:2022gbk, Borden:2023ugd} and the intercepts obtained by BER \cite{Bartels:1995iu,Bartels:1996wc}. See the Appendix of \cite{Borden:2023ugd} for an explanation of the potential origin of this discrepancy.

\section{OAM to helicity PDF ratios}
\label{sec:ratios}

Now that we have the analytic expressions for the quark and gluon OAM distributions at small-$x$ and large-$N_c$, we can explicitly calculate the ratio of the OAM distributions to the helicity PDFs. These ratios have previously been studied in \cite{Hatta:2016aoc,Hatta:2018itc,Boussarie:2019icw,More:2017zqp, Kovchegov:2019rrz, Kovchegov:2023yzd}, with a discrepancy in the quark ratio between \cite{Boussarie:2019icw} and the numerical solution of \cite{Kovchegov:2023yzd}. We seek here to investigate this claim and reveal any insight that can be obtained from the exact solution of the moment amplitude evolution. 

To start, we evaluate \eqs{exact_dists} for specific initial conditions. Then we take the ratio of \eq{eLq} to \eq{eDE} and \eq{eLG} to \eq{eDG} to obtain the quark and gluon ratios respectively. In \cite{Kovchegov:2023yzd}, it was found the asymptotic ($x \to 0$) behavior of these ratios is independent of the initial conditions of both the helicity and moment evolution. Therefore, without loss of generality, we choose the following initial conditions 
\begin{subequations}
    \label{ics_ratios}
\begin{align}
    G^{(0)}(s_{10}, \eta) &=1, \\
    \bar{I}_{3}^{(0)}(s_{10}, \eta) = \vec{I}^{(0)}(s_{10}, \eta) = G_2^{(0)}(s_{10}, \eta) &= 0.
\end{align}
\end{subequations}
Using \eqs{ics_ratios} in \eqs{aux_defs} gives
\begin{subequations}
    \label{simp_ics}
\begin{align}
    G^{(0)}_{\og} &= \frac{1}{\og}, \\ 
    \bar{I}_{3\og}^{(0)} = \vec{I}^{(0)}_{\og} = G^{(0)}_{2\og} &= 0, \\
    G_{\og} &= \frac{\gamma_\omega^-(\og - 4)}{4\gamma (\gamma- \gamma_\omega^-)(2\delta^+_\omega - \omega)},
    \\ 
    G_{2\og} &= \frac{\gamma^-_\omega}{2\gamma (\gamma- \gamma_\omega^-)(2\, \delta^+_\omega - \, \omega)}, \\ 
    \bar{I}_{3 \og} &= \frac{\gamma^-_\omega (2-\gamma \omega)(\delta^-_\omega \omega - 1)}{4\gamma \omega^2 (\gamma-\gamma^-_\omega)(2\, \delta^+_\omega - \omega)(\gamma - \bar{\gamma}^-_\omega)(\gamma - \bar{\gamma}^+_\omega)}, 
\end{align}
\end{subequations}
where we have defined 
\begin{align}
    2\gamma^2 \omega^2 - \gamma \omega(5 + 2\, \delta^-_\omega \omega) + 4 \equiv 2 \omega^2 (\gamma - \bar{\gamma}^+_\omega)(\gamma - \bar{\gamma}^-_\omega).
\end{align}
Now, using \eqs{simp_ics} in \eqs{exact_dists} gives 
\begin{subequations}
    \label{dists_ICs}
    \begin{align}
    \label{icLq}
    L_{q+\bar{q}}(y, t) &= \frac{N_f}{\as \pi^2} \iw \ig \, e^{\omega y + \gamma t}
    {
    \frac{
      \gamma_\omega^- \Big[
         \delta_\omega^- \omega \gamma - 2 \delta_\omega^- +  \omega^2 \gamma (\gamma - \bar{\gamma}^+_\omega)(\gamma- \bar{\gamma}^-_\omega)
      \Big]
    }{
     4 \omega^2 \gamma^2 (\gamma-\gamma^-_\omega)(  2 \delta^+_\omega- \omega)(\gamma - \bar{\gamma}^+_\omega)(\gamma- \bar{\gamma}^-_\omega)
    }
    },
    \\ \label{icLG}
    L_G(y, t) &= -\frac{2N_c}{\as \pi^2} \iw \ig e^{\omega y + \gamma t} \frac{
    {
    \Big[4 \delta_\omega^- \omega - \omega \gamma + 4 \omega^2(\gamma - \bar{\gamma}^+_\omega)(\gamma - \bar{\gamma}^-_\omega) \Big]
    }
    }{
        2 \gamma (\gamma - \gamma_\omega^-)(2 \delta^+_\omega - \omega) (\gamma - \bar{\gamma}^+_\omega)(\gamma - \bar{\gamma}^-_\omega)
    },
    \\ \label{icDE}
    \Delta \Sigma(y, t) &= -\frac{N_f}{\as \pi^2} \iw \ig e^{\omega y + \gamma t} 
    {
    \frac{
        \gamma_\omega^- 
    }{
        4 \gamma (\gamma - \gamma_\omega^-)( 2 \delta^+_\omega- \omega)
    }
    },
    \\ \label{icDG}
     \Delta G(y, t) &= \frac{2N_c}{\as \pi^2} \iw \ig e^{\omega y + \gamma t} 
     \frac{\gamma^-_\omega}{2 \gamma (\gamma - \gamma_\omega^-) (2 \delta^+_\omega - \omega)}
   ,
    \end{align}
\end{subequations}
where we have expressed both the OAM distributions and hPDFs with the arguments $y = \sqrt{\bas} \ln 1/x$ and $t = \sqrt{\bas} \ln Q^2/\Lambda^2$. We will use these arguments for the rest of this Section, since such variables appear more natural in our expressions as can be seen from \eqs{exact_dists}. {Note from \eqs{icDE} and (\ref{icDG}), we have $\Delta \Sigma(y, t) = - (N_f/4N_c) \Delta G(y,t)$.}

Now we close the $\gamma$-contours in \eqs{dists_ICs} to the left. (Note that $\mathrm{Re} \,\omega > \mathrm{Re} \,\gamma$ along the integration contours.) The result is 
\begin{subequations}
    \label{dists_omega}
    \begin{align}
    \label{oLq}
    L_{q+\bar{q}}(y, t) &= \frac{N_f}{\as \pi^2} \iw 
    {
    \frac{e^{\omega y}}{4 \omega^2 (2 \delta_\omega^+ - \omega)}
    }
    \Bigg[
    \frac{2 \delta_\omega^-}{(\bar{\gamma}^-_\omega)^2 \bar{\gamma}^+_\omega} - \omega^2 + \frac{\delta^-_\omega}{(\bar{\gamma}^+_\omega)^2 \bar{\gamma}^-_\omega \gamma^-_\omega }\Bigg( 2 \bar{\gamma}^+_\omega + 2 \gamma^-_\omega + 2 t\, \gamma_\omega^- \bar{\gamma}^+_\omega  - \omega \bar{\gamma}^+_\omega \gamma^-_\omega \Bigg)
    \\ \notag 
    & + 
    e^{t \gamma^-_\omega} \frac{
        \omega^2 \gamma_\omega^- (\bar{\gamma}^+_\omega - \gamma_\omega^-) (\bar{\gamma}^-_\omega - \gamma_\omega^-) + \delta_\omega^- (\omega \gamma^-_\omega - 2) 
    }{
    \gamma^-_\omega (\bar{\gamma}^+_\omega - \gamma_\omega^-) (\bar{\gamma}^-_\omega - \gamma_\omega^-)
    }
    + e^{t \bar{\gamma}^+_\omega} \frac{\delta_\omega^- \gamma_\omega^- (\omega \bar{\gamma}^+_\omega - 2)}{(\bar{\gamma}^+_\omega)^2 (\bar{\gamma}^+_\omega - \bar{\gamma}^-_\omega)(\bar{\gamma}^+_\omega- \gamma^-_\omega)}
    \\ \notag 
    & 
    +  e^{t \bar{\gamma}^-_\omega} \frac{\delta_\omega^- \gamma_\omega^- (\omega \bar{\gamma}^-_\omega - 2)}{(\bar{\gamma}^-_\omega)^2 (\bar{\gamma}^-_\omega - \bar{\gamma}^+_\omega)(\bar{\gamma}^-_\omega- \gamma^-_\omega)}
    \Bigg]
    \\ \label{oLG}
    L_G(y, t) &= -\frac{2N_c}{\as \pi^2} \iw \frac{e^{\omega y} \gamma_\omega^-}{2 (2 \delta^+_\omega - \omega)} \Bigg[
        e^{\gamma^-_\omega t} \frac{4(\gamma_\omega^-)^2 \omega^2 - \gamma^-_\omega \omega (11+ 4 \delta^-_\omega \omega) + 4 \delta_\omega^- \omega + 8 }{
        2 \gamma^-_\omega (\gamma^-_\omega - \bar{\gamma}^+_\omega)(\gamma^-_\omega - \bar{\gamma}^-_\omega)}
        -\frac{4 \delta_\omega^- \omega + 8}{
        2 \gamma_\omega^- \bar{\gamma}^+_\omega \bar{\gamma}^-_\omega}
        \\ \notag 
        &+ 
        e^{\bar{\gamma}^+_\omega t} \frac{4(\bar{\gamma}^+_\omega)^2 \omega^2 - \bar{\gamma}^+_\omega \omega (11+ 4 \delta^-_\omega \omega) + 4 \delta_\omega^- \omega + 8}{
        2 \bar{\gamma}^+_\omega(\bar{\gamma}^+_\omega - \gamma_\omega^-)(\bar{\gamma}^+_\omega - \bar{\gamma}^-_\omega)}
        + 
        e^{\bar{\gamma}^-_\omega t} \frac{4(\bar{\gamma}^-_\omega)^2 \omega^2 - \bar{\gamma}^-_\omega \omega (11+ 4 \delta^-_\omega \omega) + 4 \delta_\omega^- \omega + 8}{
        2 \bar{\gamma}^-_\omega(\bar{\gamma}^-_\omega - \gamma_\omega^-)(\bar{\gamma}^-_\omega - \bar{\gamma}^+_\omega)}
    \Bigg],
    \\ \label{oDE}
    \Delta \Sigma(y, t) &= -\frac{N_f}{\as \pi^2} \iw \frac{e^{\omega y} \omega}{4(\omega - 2\delta^+_\omega)} 
    {\Bigg[ 
    1 - e^{\gamma^-_\omega t}
    \Bigg]},
    \\ \label{oDG}
     \Delta G(y, t) &= \frac{2N_c}{\as \pi^2} \iw  \frac{e^{\omega y}}{2({2} \delta^+_\omega - \omega)} \Bigg[ e^{\gamma^-_\omega t} - 1 \Bigg]
   .
   \end{align}
\end{subequations}

Due to the complex branch structure of the integrands in \eqs{dists_omega}, we approximate the $\omega$-integrals. Since we would like to determine the OAM to hPDF ratios in the asymptotic limit, where there is little to no dependence on the initial conditions, it suffices to evaluate the integrals in \eqs{dists_omega} to leading order in $y$ only. The leading order behavior is given by the rightmost singularity in the $\omega$-plane, as mentioned above. Some of the branch structure beyond this rightmost singularity is shown in Fig. 6 of \cite{Kovchegov:2023yzd}. There is a branch cut along the real axis from $\omega_b$ to some $\omega_b^\prime$ where $\omega_b^\prime < \omega_b$. Therefore, to evaluate any one of the integrals in \eqs{dists_omega}, one can wrap the contour across this branch cut and approximate the integrals as the integrals of the discontinuity across the branch cut. If we label the integrands (including pre-factors) of \eqs{dists_omega} via $L_{q+\bar{q}, \omega}, L_{G, \omega}, \Delta \Sigma_\omega,$ and $\Delta G_\omega$, we may write this approximation as 
\begin{subequations}
    \label{xi_eqns}
    \begin{align}
        \label{xi_Lq}
        L_{q+\bar{q}}(y,t) &\approx \lim_{\epsilon \to 0^+} \int \displaylimits^\infty_0 \frac{d\xi}{2\pi i} \left(L_{q+\bar{q}, \omega_b - \xi + i \epsilon} - L_{q+\bar{q}, \omega_b - \xi - i \epsilon} \right),
        \\ 
        \label{xi_LG}
        L_{G}(y,t) &\approx \lim_{\epsilon \to 0^+} \int \displaylimits^\infty_0 \frac{d\xi}{2\pi i} \left(L_{G, \omega_b - \xi + i \epsilon} - L_{G, \omega_b - \xi - i \epsilon} \right), 
        \\
        \label{xi_DE}
         \Delta \Sigma(y,t) &\approx \lim_{\epsilon \to 0^+} \int \displaylimits^\infty_0 \frac{d\xi}{2\pi i} \left(\Delta \Sigma_{\omega_b - \xi + i \epsilon} - \Delta \Sigma_{\omega_b - \xi - i \epsilon} \right), 
         \\
        \label{xi_DG}
          \Delta G(y,t) &\approx \lim_{\epsilon \to 0^+} \int \displaylimits^\infty_0 \frac{d\xi}{2\pi i} \left(\Delta G_{\omega_b - \xi + i \epsilon} - \Delta G_{\omega_b - \xi - i \epsilon} \right), 
    \end{align}
\end{subequations}
where, due to the $y$-dependent exponentials in \eqs{dists_omega}, we have extended the $\xi$-integration to infinity, and discarded the rest of the contribution from closing the contour as subleading. One can also think of this as sending $\omega_b^\prime \to -\infty$. Now it is a simple matter to plug \eqs{dists_omega} into \eqs{xi_eqns} and compute the integrals. However, since we are only interested in the leading order (in $y$) result, one can expand the integrands (excluding the large exponentials $e^{-\xi y}$) around $\xi = 0$, and keep only the leading terms. The resulting integral is then a power series in $1/y$ multiplied by the leading exponential. For example, this series is explicitly constructed for \eqs{xi_DE} and (\ref{xi_DG}) in the Appendix of \cite{Kovchegov:2023yzd}. When one takes the ratio of \eq{xi_Lq} to \eq{xi_DE} or \eq{xi_LG} to \eq{xi_DG}, the result is again a series in $1/y$. We can then write the ratios of \eq{xi_Lq} to \eq{xi_DE} and \eq{xi_LG} to \eq{xi_LG} as
\begin{subequations}
    \label{ratios_forms}
    \begin{align}
        \frac{L_{q+\bar{q}}(y,t)}{\Delta \Sigma(y,t)} = R_q(t) + \frac{B_q(t)}{y} + \mathcal{O}\left(\frac{1}{y^2} \right),
        \\ 
        \frac{L_{G}(y,t)}{\Delta G(y,t)} = R_G(t) + \frac{B_G(t)}{y} + \mathcal{O}\left(\frac{1}{y^2} \right).
    \end{align}
\end{subequations}
The explicit expressions for the asymptotic ratios, $R_{q}(t)$ and $R_G(t)$, are given in Appendix \ref{sec:exact_ratios}. In Fig. \ref{fig:ratios}, we show $R_q(t), B_q(t), R_G(t),$ and $B_G(t)$ for the branch cut discontinuity approximation in \eqs{xi_eqns} as a function of $t$. In Fig. \ref{fig:ratios}, we plot $R_q(t), B_q(t), R_G(t), B_G(t)$ in panels (a), (b), (c), and (d) respectively. 
\begin{figure}[ht!]
     \centering
     \begin{subfigure}[b]{0.48\textwidth}
         \centering
         \includegraphics[width=\textwidth]{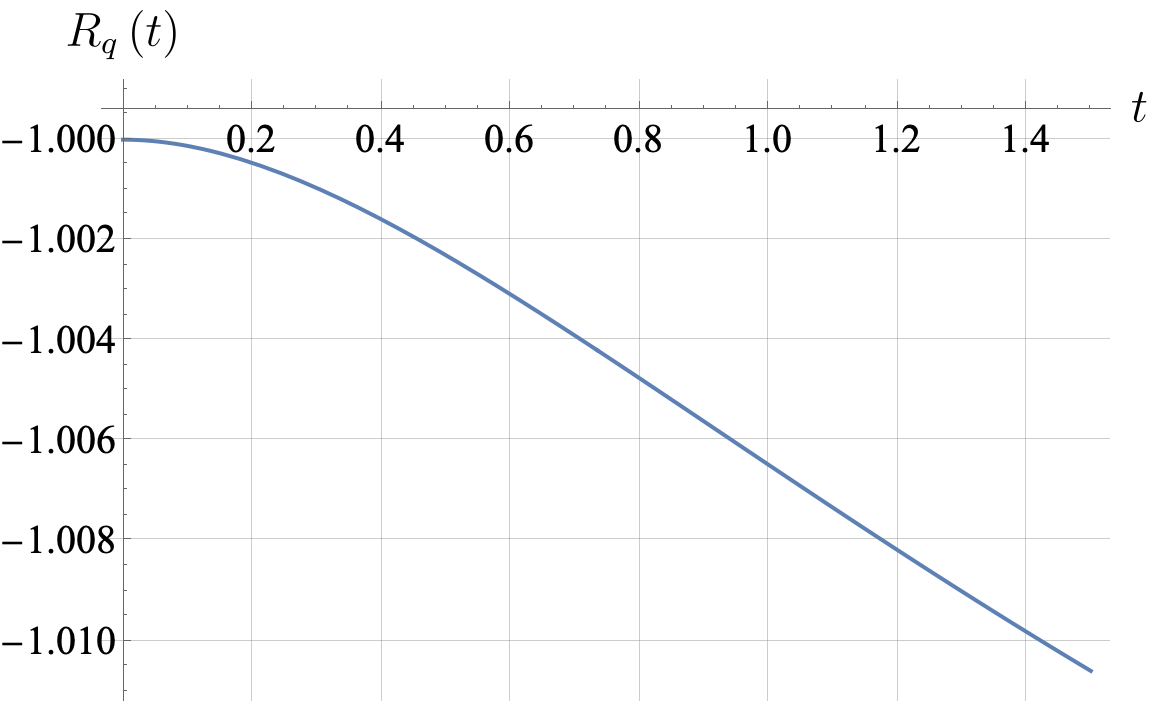}
         \caption{}
     \end{subfigure}
     \begin{subfigure}[b]{0.48\textwidth}
         \centering
         \includegraphics[width=\textwidth]{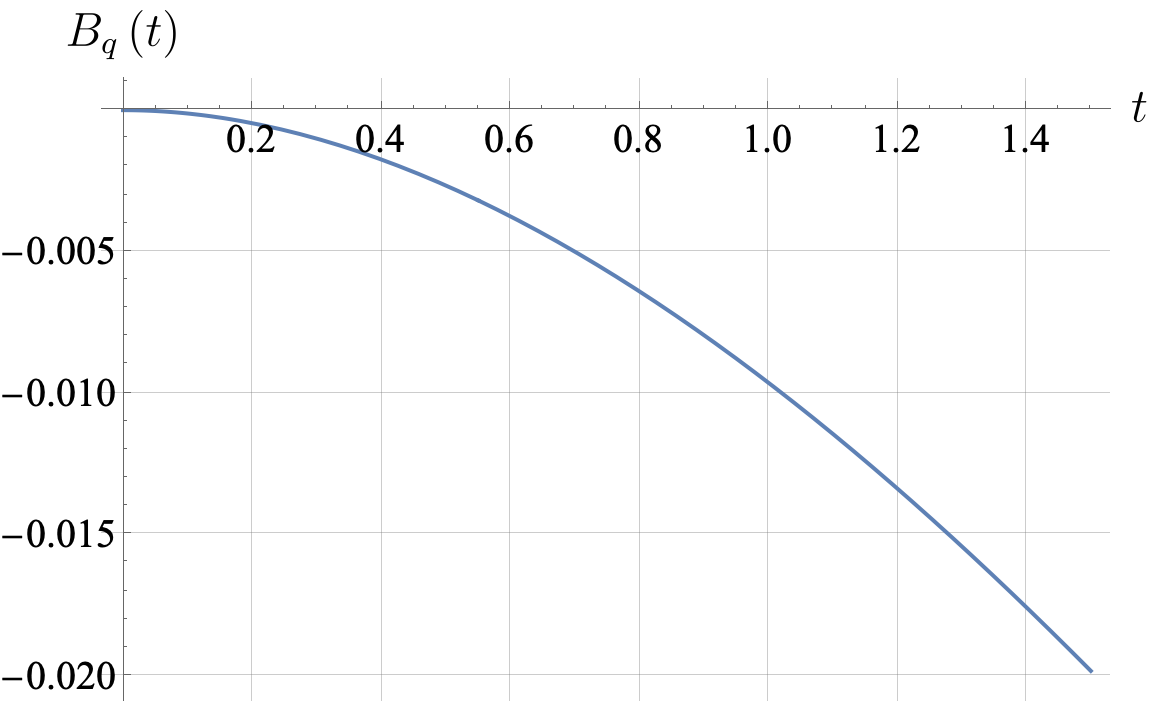}
         \caption{}
     \end{subfigure}
     \begin{subfigure}[b]{0.48\textwidth}
         \centering
         \includegraphics[width=\textwidth]{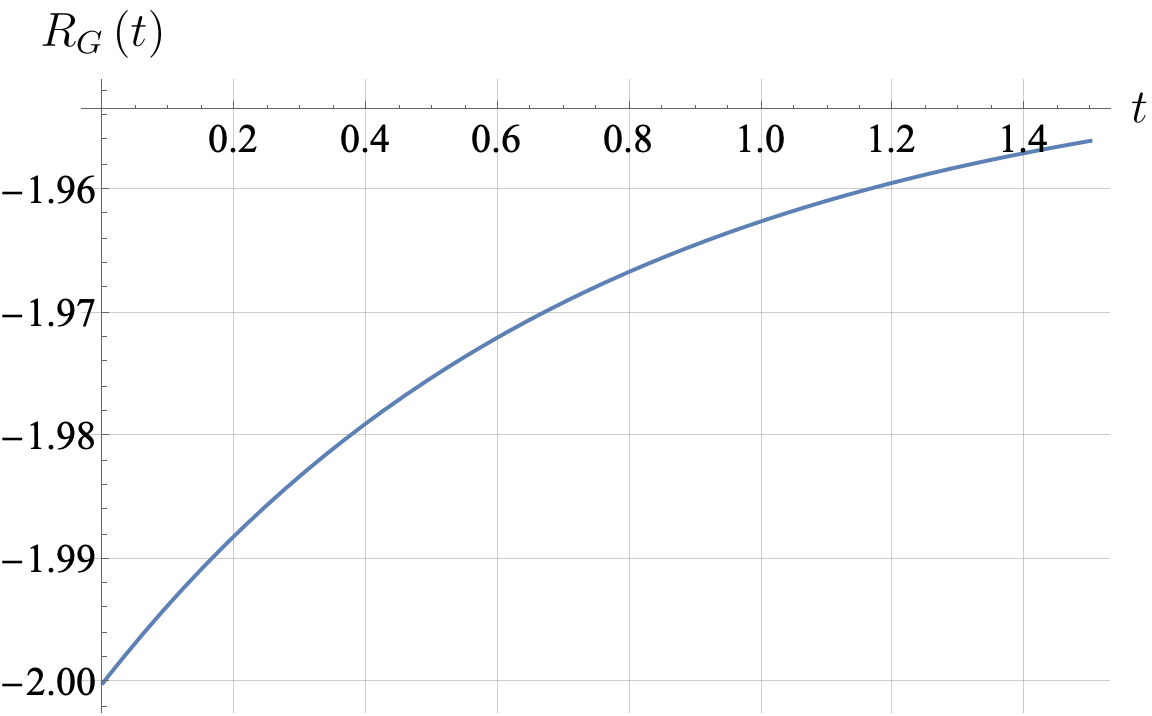}
         \caption{}
     \end{subfigure}
     \begin{subfigure}[b]{0.48\textwidth}
         \centering
         \includegraphics[width=\textwidth]{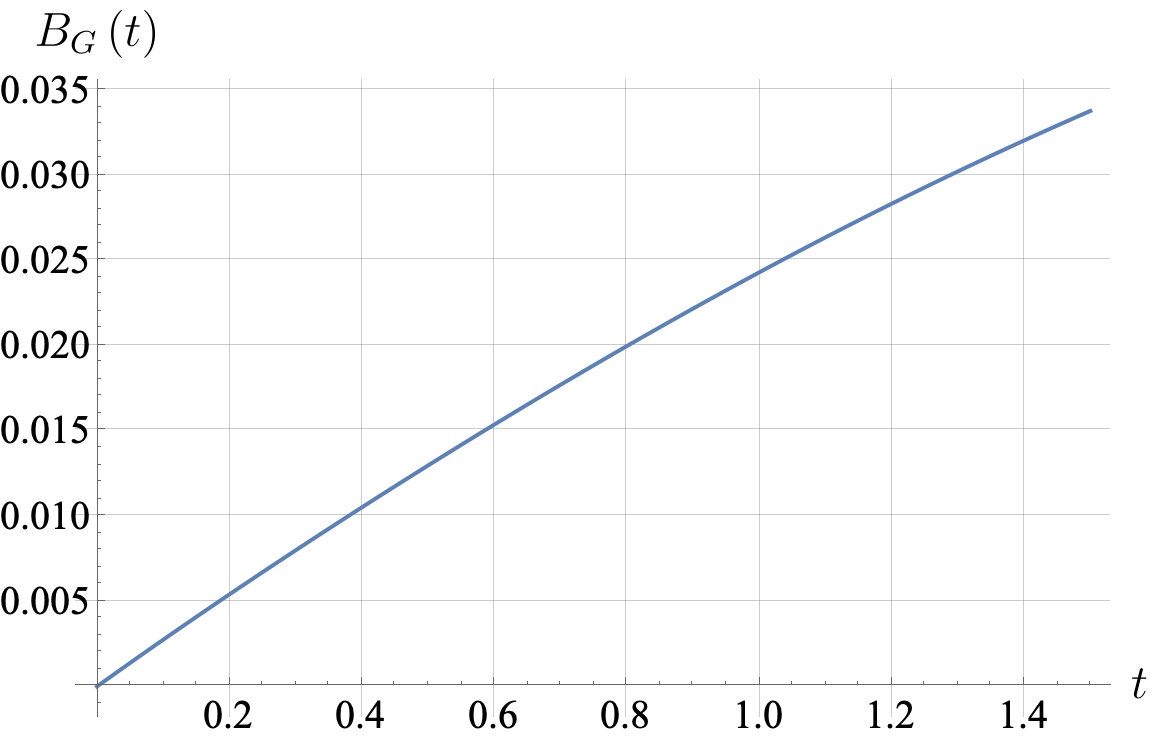}
         \caption{}
     \end{subfigure}
        \caption{Plots of the coefficients $R_q(t), B_q(t), R_G(t), B_G(t)$ from \eqs{ratios_forms} as a function of $t$ resulting from the branch cut discontinuity approximation in \eqs{xi_eqns}. Here $N_c = 3$ and $\alpha_s = 0.25$.}
        \label{fig:ratios}
\end{figure}
Additionally, we give the explicit values of $R_q(t), B_q(t), R_G(t)$, and $B_G(t)$ for some values of $t$ in Table \ref{tab:ratios}. 

\begin{table}[ht!]
    \centering
    \begin{tabular}{|c|c|c|c|c|}
        \hline
        \hspace{0.25cm} $t$ \hspace{0.25cm} & \hspace{0.25cm} $R_q\left(t \right)$ \hspace{0.25cm} &
        \hspace{0.25cm} $B_q\left(t \right)$ \hspace{0.25cm} & 
        \hspace{0.25cm} $R_G\left(t \right)$ \hspace{0.25cm} & 
        \hspace{0.25cm} $B_G\left(t \right)$ \hspace{0.25cm} \\
        \hline
        $0.1$ & { $-1.000$ }& {$0.000$} & $-1.994$ & $0.002$ \\
        $0.5$ & {$-1.002$} & {$-0.003$} & $-1.975$ & $0.013$ \\
        $1.0$ & {$-1.006$} & {$-0.010$} & $-1.962$ & $0.024$ \\
        $1.5$ & {$-1.011$} & {$-0.020$} & $-1.956$ & $0.034$ \\
        \hline
    \end{tabular}
    \caption{Specific values of the coefficients $R_q(t), B_q(t), R_G(t), B_G(t)$ in \eqs{ratios_forms} for various values of $t = \sqrt{\bas} \ln Q^2/\Lambda^2$. Here $N_c=3$ and $\alpha_s=0.25$.}
    \label{tab:ratios}
\end{table}

We should compare the results in Fig. \ref{fig:ratios} and Table \ref{tab:ratios} to the predictions from \cite{Boussarie:2019icw}. From Eqs.~(6) and (7) in \cite{Boussarie:2019icw} {(see also Eq.~(24) there)}, after using the fact that the parameter $\alpha \sim \sqrt{\as}$ is perturbatively small, we see, in the DLA, \cite{Boussarie:2019icw} predicts
\begin{subequations}\label{BHY}
\begin{align}
    \frac{L_{q+\bar{q}}(y, t)}{\Delta \Sigma(y,t)} &= -1, \label{BHYq} \\
    \frac{L_{G}(y,t)}{\Delta G(y,t)} &= -2.
\end{align}
\end{subequations}
Comparing these results to Fig. \ref{fig:ratios} and Table \ref{tab:ratios}, we see that we predict nearly the same value for the asymptotic {quark and gluon ratios. Furthermore, in} both the quark and gluon sectors, we note from Fig. \ref{fig:ratios}, our results are dependent on $t = \sqrt{\bas} \ln Q^2/\Lambda^2$, and as $t$ increases, the asymptotic ratios, $R_q(t)$ and $R_G(t)$, get further away from the results of \cite{Boussarie:2019icw} (this is true even for asymptotically large $t$, see Appendix \ref{sec:exact_ratios}). Therefore, we conclude that both the quark and gluon ratios determined here disagree with the results of \cite{Boussarie:2019icw}. This is contrast to \cite{Kovchegov:2023yzd}, where the numerically small difference in the gluon sector could have been attributed to underestimation of the numerical error. 

 The ratios in Fig. \ref{fig:ratios} are consistent with the numerical findings of \cite{Kovchegov:2023yzd}. {Other than the difference between LCOT helicity evolution and BER IREE-based helicity evolution (see \cite{Borden:2023ugd} for details), there are a few potential sources of discrepancy between the ratios of \cite{Boussarie:2019icw} and ours in Fig. \ref{fig:ratios} and Table \ref{tab:ratios}. One potential source of discrepancy is the large-$N_c$ limit we take here. For the quark ratio, our implementation of the large-$N_c$ limit is only approximately correct, since one needs to assume the existence of at least one external hard quark to calculate $\Delta \Sigma$ and $L_{q+\bar{q}}$. This hard quark line may result in soft quark emissions from evolution, which are omitted here (see the Conclusions section of \cite{Kovchegov:2020hgb} for more details) .} 
 One could hope that taking the large-$N_c\&N_f$ limit, as done in \cite{Cougoulic:2022gbk}, might account for the discrepancy. However, a preliminary numerical solution of the large-$N_c\&N_f$ moment amplitude evolution equations for $N_f/N_c \to 0$ has been constructed and has not been able to account for the discrepancy in the quark ratio. A full analysis of the large-$N_c\&N_f$ limit is left for future work. Additionally, the Wandzura-Wilczek approximation \cite{Wandzura:1977qf}, which neglects the genuine twist-3 contributions to the OAM distributions, is employed in \cite{Boussarie:2019icw}. Although the twist-3 contributions, particularly their evolution in $Q^2$, have been studied in \cite{Hatta:2019csj}, a dedicated small-$x$ study has not yet been performed. So far in this work, we have not identified the twist-3 contribution to the OAM distributions explicitly. To examine their effect on the OAM to hPDF ratios, let us attempt to isolate the twist-3 contributions explicitly.
 
We start with the following relations derived in \cite{Hatta:2012cs}, 
\begin{subequations} \label{eom_rels}
    \begin{align}
        L_{q+\bar{q}}(x,Q^2) &= x \int \displaylimits^1_x \frac{dx^\prime}{x^\prime} \Big[H_q(x^\prime, Q^2) + E_q(x^\prime, Q^2) \Big] - x \int \displaylimits^1_x \frac{dx^\prime}{x^{\prime 2}} \Delta \Sigma(x^\prime, Q^2) + L^3_{q+\bar{q}}(x,Q^2), \\ 
        L_{G}(x,Q^2) &= x \int \displaylimits^1_x \frac{dx^\prime}{x^\prime} \Big[H_g(x^\prime, Q^2) + E_g(x^\prime, Q^2) \Big] - 2 x \int \displaylimits^1_x \frac{dx^\prime}{x^{\prime 2}} \Delta G(x^\prime, Q^2) + L^3_{G}(x,Q^2),
    \end{align}
\end{subequations}
where $H_{q(g)}(x,Q^2), E_{q(g)}(x,Q^2)$ are the standard unpolarized quark (gluon) twist-2 generalized parton distributions (GPDs) at zero skewness and in the limit of zero momentum transfer. $L^3_{q+\bar{q}}(x,Q^2)$ and $L_G^3(x,Q^2)$ are the twist-3 contributions to the quark and gluon OAM distributions respectively. They can be expressed using twist-3 GPDs \cite{Hatta:2012cs}, and they are the neglected contributions in the Wandzura-Wilczek approximation. Since we are working in the DLA, we can neglect the unpolarized twist-2 GPDs, as their evolution is single-logarithmic.

Then, via \eqs{eom_rels}, we can write the DLA expressions for the twist-3 contributions to the quark and gluon OAM distributions as 
\begin{subequations} \label{dla_eoms}
    \begin{align}
        L^3_{q+\bar{q}}(x,Q^2) &= L_{q+\bar{q}}(x,Q^2) + x \int \displaylimits^1_x \frac{dx^\prime}{x^{\prime 2}} \Delta \Sigma(x^\prime, Q^2) , \\ 
        L^3_{G}(x,Q^2) &= L_{G}(x,Q^2) +  2 x \int \displaylimits^1_x \frac{dx^\prime}{x^{\prime 2}} \Delta G(x^\prime, Q^2).
    \end{align}
\end{subequations}
Similarly, we write the DLA twist-2, Wandzura-Wilczek (WW) contribution to the OAM distributions as \cite{Hatta:2012cs, Boussarie:2019icw}
\begin{subequations} \label{ww_approx}
    \begin{align}
        L^{WW}_{q+\bar{q}}(x,Q^2) &=  - x \int \displaylimits^1_x \frac{dx^\prime}{x^{\prime 2}} \Delta \Sigma(x^\prime, Q^2),
        \\
        L^{WW}_G(x,Q^2) &= - 2 x \int \displaylimits^1_x \frac{dx^\prime}{x^{\prime 2}} \Delta G(x^\prime, Q^2).
    \end{align}
\end{subequations}
Then, using \eqs{dla_eoms} and (\ref{ww_approx}), we may write the asymptotic limit of \eqs{ratios_forms} as 
\begin{subequations} \label{twist_rats}
    \begin{align} 
        \lim_{y\to \infty} \frac{L_{q+\bar{q}}(y,t)}{\Delta \Sigma(y, t)} &= \lim_{y\to \infty} \frac{L^{WW}_{q+\bar{q}}(y,t)}{\Delta \Sigma(y,t)} + \lim_{y\to \infty} \frac{L^{3}_{q+\bar{q}}(y,t)}{\Delta \Sigma(y,t)} \equiv R^{WW}_q(t) + R^3_q(t), 
        \\ 
        \lim_{y\to \infty} \frac{L_{G}(y,t)}{\Delta G(y, t)} &= \lim_{y\to \infty} \frac{L^{WW}_{G}(y,t)}{\Delta G(y,t)} + \lim_{y\to \infty} \frac{L^{3}_{G}(y,t)}{\Delta G(y,t)} \equiv R^{WW}_G(t) + R^3_G(t),
    \end{align} 
\end{subequations}
where we have again used the more convenient variables $y = \sqrt{\bas}\ln(1/x),\, t=\sqrt{\bas}\ln(Q^2/\Lambda^2)$ in the arguments of the OAM distributions and the hPDFs. 

\begin{figure}[ht!]
     \centering
     \begin{subfigure}[b]{0.6\textwidth}
         \centering
         \includegraphics[width=\textwidth]{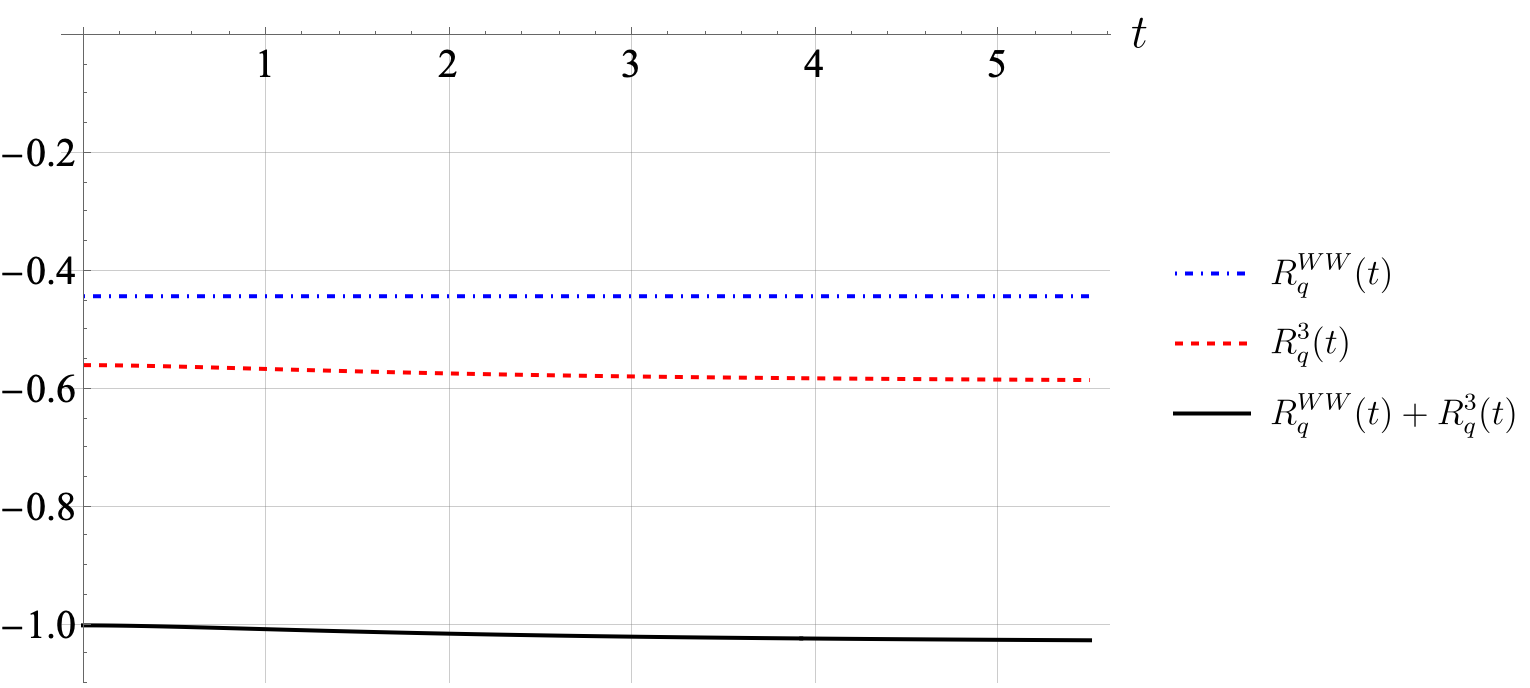}
         \caption{}
     \end{subfigure}
     \begin{subfigure}[b]{0.6\textwidth}
         \centering
    \includegraphics[width=\textwidth]{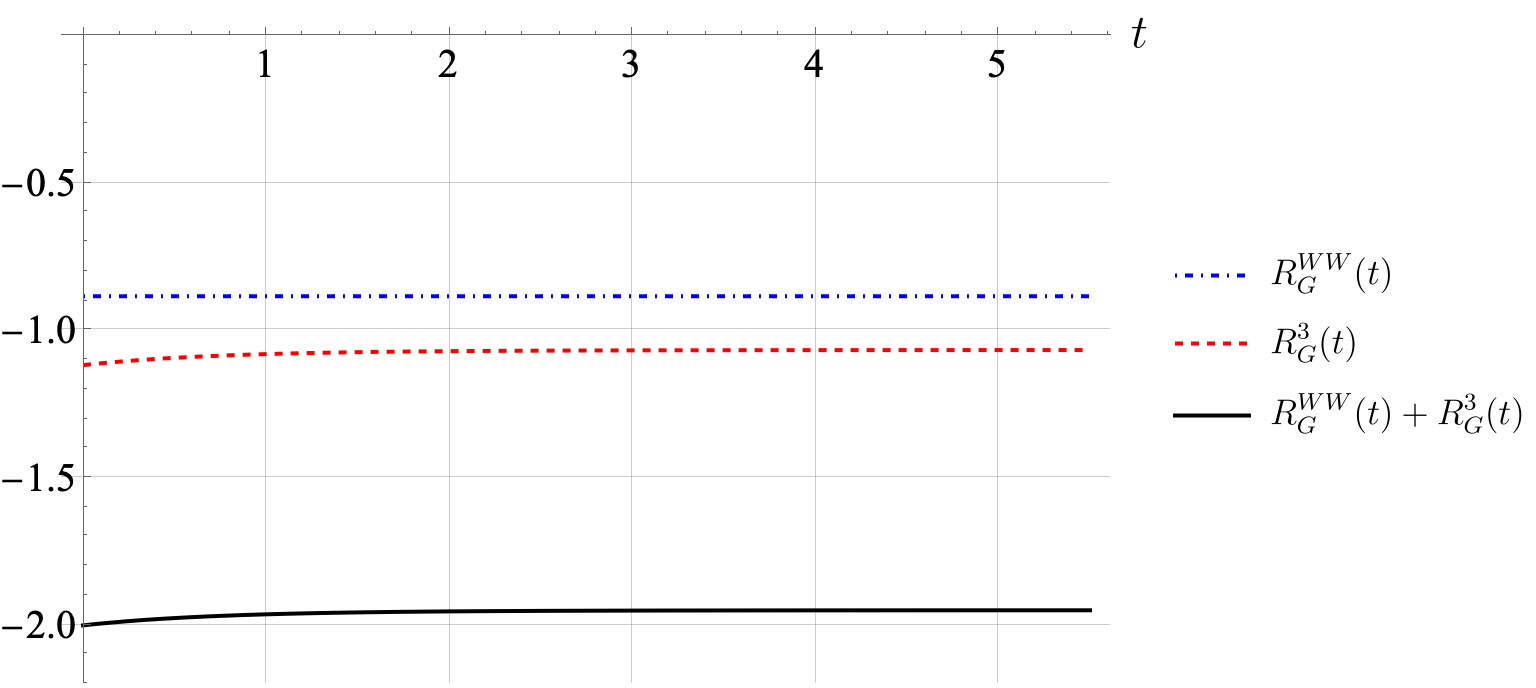}
         \caption{}
     \end{subfigure}
    \caption{Plots of the asymptotic ($x \to 0$) OAM to hPDF ratios for the twist-2 ($R^{WW}(t)$), twist-3 ($R^3(t)$) and full ($R^{WW}(t) + R^3(t)$) OAM distribution contributions in the (a) quark and (b) gluon sector as a function of $t = \sqrt{\bas} \ln(Q^2/\Lambda^2)$. Here $N_c = 3$ and $\alpha_s = 0.25$.}
        \label{fig:twist-ratios}
\end{figure}

As first observed in \cite{Boussarie:2019icw}, using the asymptotic form for helicity PDFs and the OAM distributions in \eqs{ww_approx} gives 
\begin{subequations} \label{ww_ratios}
    \begin{align}
        R^{WW}_q(t) = - \frac{1}{1+ \alpha_h}, 
        \\ 
        R^{WW}_G(t) = - \frac{2}{1+\alpha_h},
    \end{align}
\end{subequations}
where $\alpha_h = \omega_b \sqrt{\bas}$ for $\omega_b$ given above in \eq{rm_sing}. To compute the twist-3 part of the ratios in \eqs{twist_rats}, we substitute the exact expressions for the OAM distributions and the helicity PDFs from \eqs{exact_dists} into \eqs{dla_eoms}. We then perform the procedure described above to extract the OAM to hPDF ratios separately for the twist-2 (WW) and twist-3 OAM contributions. The results for the asymptotic ratios for the quark and gluon sector are shown in Fig. \ref{fig:twist-ratios}. As expected, the twist-2 contributions are exactly the values given in \eqs{ww_ratios}.\footnote{Note that we compute $R^{WW}_{q/G}$ by substituting the full DLA expressions for the helicity PDFs into \eqs{ww_approx}, not only the asymptotic forms.} We see from Fig. \ref{fig:twist-ratios} that the twist-3 contributions vary with $t$ and are similar in magnitude to the twist-2 contributions. {However, as done in \cite{Boussarie:2019icw}, it may be the case that one needs to expand in $\alpha_h$ to compute expressions in the formal DLA limit. In this case, the twist-3 contributions would be the difference between the ratios, $R_q$ and $R_G$, here and those of \cite{Boussarie:2019icw}, given in \eqs{BHY}. The resulting numerical values for the twist-3 contributions would then be roughly two orders of magnitude smaller than in Fig \ref{fig:twist-ratios}. However, above we assume no such expansion is taken.}
Therefore, according to the indirect analysis performed here, it seems as though the twist-3 contributions may be the source of the discrepancy between the OAM to hPDF ratios computed here and those computed in \cite{Boussarie:2019icw}.

\section{Conclusions and outlook}
\label{sec:conclusion}
Let us summarize what we have accomplished here. We have found an exact solution of the large-$N_c$ equations for the small-$x$ moment amplitude evolution derived in \cite{Kovchegov:2023yzd}, given in \eqs{all_oam_eqns} and (\ref{oam_neigh_eqn}). This solution is given in \eqs{the_sol} and (\ref{aux_defs}). With our solution, we have written down expressions for the quark and gluon OAM distributions at large-$N_c$ and small-$x$, given in \eqs{eLq} and (\ref{eLG}) respectively. We have determined the small-$x$ asymptotics of the quark and gluon OAM distributions in \eq{oam_asymp}. Notably, we find the small-$x$ behavior is largely driven by the mixing of the moment amplitudes $I_3, I_4, I_5, I_6$ with the dipole amplitudes $G, G_2,$ and neighbor amplitude $\Gamma_2$. Therefore, we find the discrepancy for the intercepts of the OAM distributions obtained here and the intercept obtained in \cite{Boussarie:2019icw} to be the same as the discrepancy between the helicity PDF intercepts obtained in the LCOT formalism \cite{Borden:2023ugd} and the BER formalism \cite{Bartels:1995iu,Bartels:1996wc}. 

We have also studied the ratios of the quark and gluon OAM distributions to their helicity PDF counterparts in the small-$x$ region. In \cite{Kovchegov:2023yzd}, the quark and gluon ratios in the WW approximation from \eqs{ww_ratios} were approximated by $R_q^{WW} \approx -1$ and $R_G^{WW} \approx -2$, with the numerical results of \cite{Kovchegov:2023yzd} indicating {an agreement} of the net quark ratio $R_q = R_q^{WW} + R^3_{q}$ and $R_q^{WW}$ as well as an agreement between the net ratio gluon ratio $R_G = R_G^{WW} + R^3_G$ and $R_G^{WW}$ within the accuracy of the numerical approximation. {We have found small, but non-zero disagreement between $R_G$ and $R^{WW}_G$ as well as $R_q$ and $R^{WW}_q$.} We have tentatively traced these discrepancies to the twist-3 contributions to the OAM distributions. However, a more complete study is warranted to verify this conclusion. 

\section*{Acknowledgments}

The author would like to thank Yuri Kovchegov,  Jeremy Borden and Yoshitaka Hatta for extensive advice and helpful discussions. 

This material is based upon work supported by the U.S. Department of Energy, Office of Science, Office of Nuclear Physics under Award Number DE-SC0004286 and is performed within the framework of the Saturated Glue (SURGE) Topical Theory Collaboration.


\appendix 


\section{Boundary conditions} 
\label{sec:BC}

From \eqs{vec_oam_eqns} we see there are several boundary conditions that need to be satisfied. From \eqs{sep_vec_eqn} and (\ref{sep_neigh_eqn}), we need
\begin{subequations}
    \label{bcs_1}
    \begin{align}
        \vec{I}(s_{10} = 0, \eta) &= 
        \vec{I}^{(0)}(s_{10} = 0, \eta), 
        \\ 
        \vec{I}(s_{10}, \eta = s_{10}) &= 
        \vec{I}^{(0)}(s_{10}, \eta = s_{10}),
        \\ 
        \label{bcgm1}
        \vec{\Gamma}(s_{10} = 0, s_{21}, \eta^\prime) &= 
        \vec{I}^{(0)}(s_{10} = 0, \eta^\prime),
        \\ 
        \label{bcgm2}
        \vec{\Gamma}(s_{10}, s_{21}, \eta^\prime = s_{21}) &= 
        \vec{I}^{(0)}(s_{10}, \eta^\prime = s_{21}).
    \end{align}
\end{subequations}
Plugging in \eqs{dm_ansatze}, we get 
\begin{subequations}
    \label{bcs}
    \begin{align}
        \iw\ig e^{\omega \eta}
        \vec{I}_{\og} &= 
        \iw\ig e^{\omega \eta}
        \vec{I}^{(0)}_{\og}, 
        \\ 
        \iw \ig e^{\gamma s_{10}}
       \vec{I}_{\og} &= 
        \iw\ig e^{\gamma s_{10}}
        \vec{I}^{(0)}_{\og}.
    \end{align}
\end{subequations}
Note \eqs{bcgm1} and (\ref{bcgm2}) are equivalent to \eqs{bcs}. Using the fact that $\bar{I}_{3\omega\gamma}, \vec{I}_{\og}, G_{\og}, G_{2\og} \to 0$ as $\omega \to \infty$ or $\gamma \to \infty$ in order for the double Laplace transforms to exist, we see that 
\begin{subequations}
    \label{van_cond}
\begin{align}
    \iw \frac{1}{\omega \gamma}
    \Big[
        \bar{I}_{3\og} \vec{V}_{\bar{I}_3} +  G_{\og} \vec{V}_G +  G_{2\og} \vec{V}_{G_2}+ M_{\mathrm{IR}} \vec{I}_{\og}
    \Big] = 
     \ig \frac{1}{\omega \gamma}
    \Big[
        \bar{I}_{3\og} \vec{V}_{\bar{I}_3} +  G_{\og} \vec{V}_G +  G_{2\og}\vec{V}_{G_2} + M_{\mathrm{IR}} \vec{I}_{\og}
    \Big] = 
     \vec{0},
\end{align}
\end{subequations}
since we may we close either the $\omega$- or the $\gamma$-contour to the right. Therefore, we see \eqs{bcs_1} are satisfied by use of \eq{i4i5_eqn} and \eqs{van_cond}.

\section{Constraining a double Laplace image and the homogeneous neighbor functions}
\label{sec:constraints}

Although we have solved the partial differential equation for $\bar{\Gamma}_3(s_{10}, s_{21}, \eta^\prime)$, \eq{gm3_pde}, we need to make sure that any constraints from the integral equation \eq{gamma3_eqn} are satisfied. To that end, we plug \eqs{gm3_res}, (\ref{gm2_sol}) and (\ref{gmI_solns}) into \eq{gamma3_eqn}. Doing the $s_{32}$ and $\eta^{\dprime}$ integrals, we arrive at the following constraint
\begin{align}
    0 &= \iw\ig 
         e^{\omega(\eta^\prime - s_{10}) + \gamma s_{10}} \left(\ithreeO -  G_{2\og}^{(0)} + \vec{V}_{\mathrm{UV}} \cdot \vec{I}^{(0)}_{\og} \right) 
         \\ \notag 
        & - \iw \ig 
          e^{\omega(\eta^\prime- s_{21}) + \gamma s_{10}} \Bigg[ G_{2\og} - G^{(0)}_{2\og}  -\vec{V}_{\mathrm{UV}} \cdot \Big(\vec{I}_{\og} - \vec{I}^{(0)}_{\og}\Big) \Bigg]
    \\ \notag 
    & +  \int \frac{d\omega}{2\pi i} \Bigg\{
            \bar{\Gamma}^+_{3\omega}(s_{10}) \left[ 
                \frac{e^{s_{10} \delta^+_\omega}}{\omega \,\delta^+_\omega} - e^{\delta^+_\omega \eta^\prime} - \frac{e^{\omega(\eta^\prime- s_{21}) + \delta^+_\omega s_{10}}}{\omega \, \delta^+_\omega}
            \right]
    + \bar{\Gamma}^-_{3\omega}(s_{10}) \left[ 
                \frac{e^{\delta^-_\omega s_{10}}}{\omega \, \delta^-_\omega} - e^{\delta^-_\omega \eta^\prime} - \frac{e^{\omega(\eta^\prime- s_{21}) + s_{10} \delta^-_\omega}}{\omega \, \delta^-_\omega}
            \right] \Bigg\},
\end{align}
where we have made use of the following identities: $\delta^+_\omega \delta^-_\omega = 1, \delta^+_\omega + \delta^-_\omega = \omega, (\delta^{\pm}_\omega)^2 - \omega \delta^{\pm}_\omega + 1=0$. Now we do the forward Laplace transform over $\eta^\prime$ to see
\begin{align}
    \label{intconst}
    0 &= \ig 
         e^{-\omega s_{10} + \gamma s_{10}} \left(\ithreeO -  G_{2\og}^{(0)} + \vec{V}_{\mathrm{UV}} \cdot \vec{I}^{(0)}_{\og} \right) 
         \\ \notag 
        & - \ig 
          e^{-\omega s_{21} + \gamma s_{10}} \Bigg[ G_{2\og} - G^{(0)}_{2\og}  -\vec{V}_{\mathrm{UV}} \cdot \Big(\vec{I}_{\og} - \vec{I}^{(0)}_{\og}\Big) \Bigg] 
        \\ \notag &
          - \bar{\Gamma}^+_{3\omega}(s_{10})\frac{e^{-\omega s_{21} + \delta^+_\omega s_{10}}}{\omega \, \delta^+_\omega} - \bar{\Gamma}^-_{3\omega}(s_{10})\frac{e^{-\omega s_{21} + \delta^-_\omega s_{10}}}{\omega \, \delta^-_\omega}  
    \\ \notag
   & + \int \frac{d \omega^\prime}{2\pi i} \left(\frac{\bar{\Gamma}^+_{3\omega^\prime}(s_{10})}{\delta^+_{\omega^\prime}- \omega} + \frac{\bar{\Gamma}^-_{3\omega^\prime}(s_{10})}{\delta^-_{\omega^\prime} - \omega} \right)
    + \frac{1}{\omega} \int \frac{d \omega^\prime}{2\pi i} 
    \left(
        \bar{\Gamma}^+_{3\omega^\prime}(s_{10}) \frac{e^{\delta^+_{\omega^\prime} s_{10}}}{\omega^\prime \, \delta^+_{\omega^\prime}} 
        +
        \bar{\Gamma}^-_{3\omega^\prime}(s_{10}) \frac{e^{\delta^-_{\omega^\prime} s_{10}}}{\omega^\prime \, \delta^-_{\omega^\prime}}
    \right).
\end{align}
The second and third lines of \eq{intconst} are dependent on $s_{21}$, while the first and fourth lines are independent of $s_{21}$. Since \eq{intconst} is valid for $s_{21} >0$, we conclude the second and third lines as well as the first and fourth lines must sum to zero separately. Therefore, we have
\begin{subequations}
    \label{first_consts}
\begin{align}
    \label{constG_1}
     \ig 
          e^{\gamma s_{10}} \Bigg[ G_{2\og} - G^{(0)}_{2\og}  -\vec{V}_{\mathrm{UV}} \cdot \Big(\vec{I}_{\og} - \vec{I}^{(0)}_{\og}\Big) \Bigg] 
    + \bar{\Gamma}^+_{3\omega}(s_{10})\frac{e^{\delta^+_\omega s_{10}}}{\omega \, \delta^+_\omega} + \bar{\Gamma}^-_{3\omega}(s_{10})\frac{e^{\delta^-_\omega s_{10}}}{\omega \, \delta^-_\omega} &= 0, 
    \\ \label{last_const}
   \ig 
    e^{-\omega s_{10} + \gamma s_{10}} \left(\ithreeO -  G_{2\og}^{(0)} + \vec{V}_{\mathrm{UV}} \cdot \vec{I}^{(0)}_{\og} \right)   + \int \frac{d \omega^\prime}{2\pi i} \left(\frac{\bar{\Gamma}^+_{3\omega^\prime}(s_{10})}{\delta^+_{\omega^\prime}- \omega} + \frac{\bar{\Gamma}^-_{3\omega^\prime}(s_{10})}{\delta^-_{\omega^\prime} - \omega} \right)&
    \\ \notag 
   \hspace{2cm} + \frac{1}{\omega} \int \frac{d \omega^\prime}{2\pi i} 
    \left(
        \bar{\Gamma}^+_{3\omega^\prime}(s_{10}) \frac{e^{\delta^+_{\omega^\prime} s_{10}}}{\omega^\prime \, \delta^+_{\omega^\prime}} 
        +
        \bar{\Gamma}^-_{3\omega^\prime}(s_{10}) \frac{e^{\delta^-_{\omega^\prime} s_{10}}}{\omega^\prime \, \delta^-_{\omega^\prime}}
    \right) &= 0.
\end{align}
\end{subequations}
To further constrain the functions $\bar{\Gamma}_{3\omega}^\pm(s_{10})$ and the double Laplace image $\bar{I}_{3\og}$, we note from \eq{gamma3_eqn} we need to satisfy the following boundary condition
\begin{align}
    \label{gm_rel}
    \bar{\Gamma}_3(s_{10}, s_{21} = s_{10}, \eta^\prime) = \bar{I}_3 (s_{10}, \eta^\prime).
\end{align}
Substituting \eqs{dm_ansatze} and (\ref{gm3_res}) into \eq{gm_rel}, and doing the forward Laplace transform over $\eta-s_{10}$, we get 
\begin{align}
    \label{constG_2}
    \bar{\Gamma}^+_{3\omega}(s_{10}) \, e^{\delta^+_\omega s_{10}} + \bar{\Gamma}^-_{3\omega}(s_{10}) \, e^{\delta^-_\omega s_{10}} + \ig e^{ \gamma s_{10}} \left( G_{2\og} - \vec{V}_{\mathrm{UV}} \cdot \vec{I}_{\og}  - \bar{I}_{3\omega \gamma} \right) = 0.
\end{align}
With \eqs{constG_2} and (\ref{constG_1}), we can solve for $\bar{\Gamma}^{\pm}_{3\omega}(s_{10})$. We find
\begin{align}
    \label{appenB_gmres}
    \bar{\Gamma}^{\pm}_{3\omega}(s_{10}) &= e^{- \delta^{\pm}_\omega s_{10}} \frac{\delta^\pm_{\omega}}{\delta^\pm_\omega - \delta^\mp_\omega} \ig e^{\gamma s_{10}}
    \Bigg\{(\omega \delta^\mp_\omega -1) \Big[ G_{2\og} - G_{2\og}^{(0)} - \vec{V}_{\mathrm{UV}} \cdot \left(\vec{I}_{\og} - \vec{I}_{\og}^{(0)} \right) \Big] 
    \\ \notag 
    & \hspace{10cm}
    - G_{2\og}^{(0)} + \vec{V}_{\mathrm{UV}} \cdot \vec{I}^{(0)}_{\og} + \ithree
    \Bigg\},
\end{align}
which is exactly \eq{gm_funcs} in the main text.

Now we only have to satisfy \eq{last_const}. Using the fact that $G_{2\og}, \bar{I}_{3\og}, \vec{I}_{\og}, \bar{I}_{3\og}^{(0)}, G_{2\og}^{(0)}, \vec{I}^{(0)}_{\og} \to 0$ as $\omega \to \infty$, we conclude that $\bar{\Gamma}^\pm_{3\omega}(s_{10}) \,e^{\delta^\pm_\omega s_{10}} \to 0$ when $\omega \to \infty$. This fact allows us to close the contour to the right in the last term of \eq{last_const}. We are left with 
\begin{align}
    \ig e^{(\gamma - \omega)s_{10}} \left( \ithreeO - G^{(0)}_{2\og} + \vec{V}_{\mathrm{UV}} \cdot \vec{I}^{(0)}_{\og} \right)  &=  \int \frac{d \omega^\prime}{2\pi i} \left(\frac{\bar{\Gamma}^+_{3\omega^\prime}(s_{10})}{\omega- \delta^+_{\omega^\prime}} + \frac{\bar{\Gamma}^-_{3\omega^\prime}(s_{10})}{\omega- \delta^-_{\omega^\prime}} \right) 
    \\ \notag 
    &= -\int \frac{d \omega^\prime}{2\pi i} \left(\frac{\omega - \delta^-_{\omega^\prime}}{\omega}\frac{\bar{\Gamma}^+_{3\omega^\prime}(s_{10})}{\omega^\prime - \left(\omega + \frac{1}{\omega}\right)} + \frac{\omega - \delta^+_{\omega^\prime}}{\omega}\frac{\bar{\Gamma}^-_{3\omega^\prime}(s_{10})}{\omega^\prime - \left(\omega + \frac{1}{\omega}\right)} \right).
\end{align}
Now we close the contour to the right, picking up the pole at $\omega + \frac{1}{\omega}$. We are left with 
\begin{align}
    \label{appB_1}
    \ig e^{\gamma s_{10}} \left(\bar{I}^{(0)}_{3\omega \gamma} - G^{(0)}_{2\og} + \vec{V}_{\mathrm{UV}} \cdot \vec{I}^{(0)}_{\og} \right) = \left( 1- \frac{1}{\omega^2}\right) \bar{\Gamma}^+_{3,\omega+\frac{1}{\omega}}(s_{10})\, e^{\omega s_{10}}.
\end{align}
Using $\delta^+_{\omega+ \frac{1}{\omega}}= \omega$, we can rewrite \eq{appB_1} as 
\begin{align}
    \ig e^{\gamma s_{10}} \left(\bar{I}^{(0)}_{3 \delta^+_{\omega + \frac{1}{\omega}} \gamma} - G^{(0)}_{2 \delta^+_{\omega + \frac{1}{\omega}} \gamma} + \vec{V}_{\mathrm{UV}} \cdot \vec{I}^{(0)}_{\delta^+_{\omega + \frac{1}{\omega}} \gamma} \right) = \left( 1- \frac{1}{\Big[\delta^+_{\omega + \frac{1}{\omega}}\Big]^2}\right) \bar{\Gamma}^+_{3,\omega+\frac{1}{\omega}}(s_{10}) \,e^{\delta^+_{\omega + \frac{1}{\omega}} s_{10}},
\end{align}
and, upon replacing $\omega + \frac{1}{\omega} \to \omega$, we have 
\begin{align}
    \ig e^{\gamma s_{10}} \left(\bar{I}^{(0)}_{3 \delta^+_{\omega} \gamma} - G^{(0)}_{2 \delta^+_{\omega} \gamma} + \vec{V}_{\mathrm{UV}} \cdot \vec{I}^{(0)}_{\delta^+_{\omega} \gamma} \right) = \left( 1- \frac{1}{\Big[\delta^+_{\omega}\Big]^2}\right) \bar{\Gamma}^+_{3,\omega}(s_{10}) \,e^{\delta^+_{\omega} s_{10}}.
\end{align}
Employing \eq{appenB_gmres}, and doing the forward Laplace transform over $s_{10}$, we find
\begin{align}
    \bar{I}^{(0)}_{3 \delta^+_{\omega} \gamma} - G^{(0)}_{2 \delta^+_{\omega} \gamma} + \vec{V}_{\mathrm{UV}} \cdot \vec{I}^{(0)}_{\delta^+_{\omega} \gamma} &= (\omega \delta^\mp_\omega -1) \Big[ G_{2\og} - G_{2\og}^{(0)} - \vec{V}_{\mathrm{UV}} \cdot \left(\vec{I}_{\og} - \vec{I}_{\og}^{(0)} \right) \Big] 
    - G_{2\og}^{(0)} + \vec{V}_{\mathrm{UV}} \cdot \vec{I}^{(0)}_{\og} + \ithree.
\end{align}
Via \eqs{i4i5_eqn}, we can solve for $\ithree$ explicitly. We get 
\begin{align}
\bar{I}_{3\og} &= 
    \frac{\vec{V}_{\mathrm{UV}} \cdot \Big(\vec{I}^{(0)}_{\og} - \vec{I}^{(0)}_{\delta^+_\omega \gamma} \Big)- G^{(0)}_{2\og} + G^{(0)}_{2 \delta^+_\omega \gamma} - \bar{I}^{(0)}_{3\delta^+_\omega \gamma}}{(\omega \delta^-_\omega -1) \vec{V}_{\mathrm{UV}} \cdot (2\og - M_{\mathrm{IR}})^{-1} \vec{V}_{\bar{I}_3} -1} 
    \\ \notag 
   & +  \frac{(\omega \delta^-_\omega -1)\Big[
        G_{2\og} - G^{(0)}_{2\og} + \vec{V}_{\mathrm{UV}} \cdot \vec{I}^{(0)}_{\og} - \vec{V}_{\mathrm{UV}} \cdot (2 \og - M_{\mathrm{IR}})^{-1}(2 \og \vec{I}^{(0)}_{\og} + G_{\og} \vec{V}_{G} + G_{2\og} \vec{V}_{G_2})
    \Big] }{(\omega \delta^-_\omega -1) \vec{V}_{\mathrm{UV}} \cdot (2\og - M_{\mathrm{IR}})^{-1} \vec{V}_{\bar{I}_3} -1},
\end{align}
which is exactly \eq{i3_exp} in the main text. 

\section{Exact expressions for the asymptotic OAM to helicity PDF ratios}
\label{sec:exact_ratios}

Here we give the exact expressions for the asymptotic ratios, $R_q(t), R_G(t)$, in \eqs{ratios_forms} using the branch cut discontinuity approximation outlined in the main text. Let us start with the quark OAM to quark hPDF ratio. Using the procedure to obtain $R_q(t)$ outlined in the main text, we find 
\begin{align}
    R_q(t) = {\frac{C_q}{t}} \Bigg\{
        p_1(\omega_b) + t \,p_2(\omega_b) &+ e^{-t \frac{\omega_b}{2}} p_3(\omega_b) 
        \\ \notag & 
        + e^{-\frac{1}{64 \omega_b} t \,(\omega_b^4 + 16 \omega_b^2 - 80)} \left[ 
            p_4(\omega_b) \cosh\left(\frac{\Omega}{16 \omega_b} t \right) 
           - p_5(\omega_b) \sinh\left(\frac{\Omega}{16 \omega_b} t \right) 
        \right]
    \Bigg \},
\end{align}
where we have used $\omega_b$ as defined in \eq{rm_sing}, and defined
\begin{subequations}
\begin{align}
    C_q &= {
        \frac{(\wb^8 - 256 \wb^4 + + 1024 \wb^2 + 8192)^{-1}}{8 \,\Omega \wb (    \wb^{14} + 64 \,\wb^{12} - 336\, \wb^{10} - 9088 \,\wb^8 + 36864\, \wb^6 + 294912\, \wb^4  - 1179648 \,\wb^2 + 1048576)^2}
    }
    \\ 
    \Omega &= \sqrt{22 \,\omega_b^4 - 2\, \omega_b^6 + 96\, \omega_b^2 - 112},
\end{align}
\end{subequations}
and also the following polynomials
\begin{subequations}
    \begin{align}
        p_1(\omega_b) &=  {
            512\,  \Omega (\wb^8 + 64 \,\wb^6 - 256 \,\wb^4 - 4096 \,\wb^2 + 8192)^2 }
        \\ \notag 
        & \hspace{1cm} 
          {\times (\wb^{14} + 8 \,\wb^{12} - 96 \,\wb^{10} - 1536\,\wb^8 - 24576 \,\wb^6 + 360448\, \wb^4 + 786432 \,\wb^2 - 4194304),}
        \\ 
        p_2(\omega_b) &= {-256 \, \Omega \wb^3 (\wb^8 + 64 \,\wb^6 - 256 \,\wb^4 - 4096 \,\wb^2 + 8192)^2}
        \\ \notag 
        & \hspace{1cm} 
        {\times (4\, \wb^{14} - 39 \,\wb^{12} - 804 \,\wb^{10} + 13664 \,\wb^8 - 63488 \,\wb^6 + 12288\, \wb^4 + 1376256 \,\wb^2 - 3407872),}
        \\
        p_3(\omega_b) &= {- \Omega (\wb^{10} - 16\, \wb^8 - 256 \,\wb^6 + 5120 \,\wb^4 - 8192 \,\wb^2 - 131072)}
        \\ \notag 
        & \hspace{1cm} 
        {\times ( \wb^{14} + 64 \,\wb^{12} - 336 \,\wb^{10} - 9088 \,\wb^8 + 36864  \,\wb^6 + 294912 \, \wb^4 -1179648 \,\wb^2 + 1048576)^2,}
        \\
        p_4(\omega_b) &= {-65536 \, \Omega \wb^4 (\wb^6 - 80 \,\wb^2 + 128)^2}
        \\ \notag 
        & \hspace{1cm} 
        {\times ( \wb^{14}  - 352 \,\wb^{10} + 3584 \,\wb^8 -23552 \,\wb^6 + 557056\, \wb^4 -6946816 \,\wb^2 +32505856 ),}
        \\
        p_5(\omega_b) &= {-262144  \,  \wb^4 (\wb^6 - 80 \,\wb^2 + 128)^2}
        \\ \notag 
        & 
        {\times (4\, \wb^{16} + 33\,  \wb^{14} - 3008 \wb^{12} + 11744 \,\wb^{10} + 426496 \,\wb^8 - 3402752 \,\wb^6 - 3833356 \, \wb^4 + 60686336 \,\wb^2 - 22020096 ).}
    \end{align}
\end{subequations}

Similar to the quark case above, following the procedure from the main text, one obtains the asymptotic gluon OAM to gluon hPDF ratio as
\begin{align}
    R_G(t) = \frac{C_G}{t} \Bigg\{
        q_1(\omega_b) + t \,q_2(\omega_b)
        + e^{-\frac{1}{64 \omega_b} t \,(\omega_b^4 + 16 \omega_b^2 - 80)} \left[ 
            q_3(\omega_b) \sinh\left(\frac{\Omega}{16 \omega_b} t \right) 
            - q_1(\omega_b) \cosh\left(\frac{\Omega}{16 \omega_b} t \right) 
        \right]
    \Bigg \},
\end{align}

where we have defined 
\begin{align}
    C_G &= -\frac{32}{\Omega 
    \left( 3\, \wb^8 - 24\, \wb^6 + 200 \,\wb^4 - 640\, \wb^2 + 512 
    \right)^2
    \left(\omega_b^2 + 8  \right)
    }, 
\end{align}
and also the following polynomials
\begin{subequations}
    \begin{align}
        q_1(\omega_b) &=  4\, \Omega \,\wb \Big( 
        \wb^{14} -7 \,\wb^{12} -251 \,\wb^{10} 
        \\ \notag & \hspace{2cm}
         +3178 \,\wb^8 
         -16264\, \wb^6 +53184  \, \wb^4  -10803 \,\wb^2 + 77824
        \Big), 
        \\ 
        q_2(\omega_b) &= \Omega  \Big( 
        4\, \wb^{18} - 45 \,\wb^{16} -594 \,\wb^{14} +12633\,\wb^{12} 
        \\ \notag & \hspace{2cm}
         -95238 \,\wb^{10} 
         +414240  \,\wb^8 
         -1114880\, \wb^6 +1740800 \, \wb^4 -1368064 \,\wb^2 + 393216
        \Big), 
        \\
        q_3(\omega_b) &=  8\, \omega_b\, (7 \,\wb^{16} -56 \,\wb^{14} -731   \,\wb^{12} 
        \\ \notag & \hspace{2cm}
         +774  \,\wb^{10} 
         +54940   \,\wb^8 
         -224848\, \wb^6 +208768  \, \wb^4 -121856\,\wb^2 + 188416
        \Big). 
    \end{align}
\end{subequations}

For large $t$, one can show the quark and gluon ratios are constants, given by 
\begin{subequations}
    \begin{align}
        R_q(t) &\approx C_q p_2(\omega_b) \approx -1.033,  \\
        R_G(t) &\approx C_G\, q_2(\omega_b) \approx -1.949.
    \end{align}
\end{subequations}

\providecommand{\href}[2]{#2}\begingroup\raggedright\endgroup

\end{document}